\documentclass{emulateapj}
\pdfoutput=1
\usepackage{lscape}
\journalinfo{Accepted for publication in AJ}
\newcommand{\HII}{H\,{\sc ii}}
\newcommand{\HST}{\textit{HST}}
\newcommand{\Spitzer}{\textit{Spitzer}}

\shorttitle {Youngest \HII\ Regions in M82 at 7~mm} \shortauthors{Tsai et al.}

\begin{document}

\title{Locating the Youngest \HII\ Regions in M82 with 7~mm Continuum Maps}

\author{Chao-Wei Tsai\altaffilmark{1}, Jean L. Turner\altaffilmark{1},
Sara C. Beck\altaffilmark{2}, David S. Meier\altaffilmark{3,4}, \& Paul
T. P. Ho\altaffilmark{5,6}}

\altaffiltext{1}{Department of Physics and Astronomy, UCLA, Los
Angeles, CA 90095-1547; email: cwtsai@astro.ucla.edu,
turner@astro.ucla.edu}
\altaffiltext{2}{Department of Physics and
Astronomy, Tel Aviv University, Ramat Aviv, Israel; email:
sara@wise1.tau.ac.il}
\altaffiltext{3}{Jansky Fellow: National Radio
Astronomy Observatory, P. O. Box 0, Socorro, NM 87801}
\altaffiltext{4}{Current address: Department of Physics, New Mexico 
Institute of Mining Technology, Socorro, NM 87801; email: dmeier@nmt.edu}
\altaffiltext{5}{Harvard-Smithsonian Center for
Astrophysics, Cambridge MA 02138; email: ho@cfa.harvard.edu}
\altaffiltext{6}{Institute of Astronomy and Astrophysics, Academia
Sinica, PO Box 23-141, Taipei 106, Taiwan}

\begin{abstract}

We present 7~mm Very Large Array continuum images of
the starburst galaxy M82. On arcsecond scales,
two-thirds of the 7mm continuum consists of free-free emission from
\HII\ regions.
In the subarcsecond resolution map, we identify 14 compact
sources, including 9 bright \HII\ regions with $N_{Lyc} >
10^{51}~sec^{-1}$. Four of the \HII\ regions have rising spectra,
implying emission measures $>10^8$~cm$^{-6}$~pc. Except for one
compact source with peculiar features, all other compact radio
sources are found in dust lanes and do not have optical 
or near-infrared continuum
counterparts. Four regions of extended, high brightness
(EM~$>10^7~{\rm cm^{-6}\,pc}$) radio emission are found in our high
resolution map, including some as large as $\sim$2\arcsec, or 30 pc,
representing either associations of small \HII\ regions, or sheetlike
structures of denser gas.
The good correlation between 7~mm emission and
\Spitzer\ IRAC $8~\micron$ continuum-removed PAH feature suggests that
PAH emission may track the recently formed OB stars. We
find an excellent correlation between molecular gas and star
formation, particularly dense gas traced by HCN, down to the $\sim$
45~pc scale in M82. 
We also find star formation efficiencies (SFEs) of 1--10\% on the same
scale, based on CO maps. The highest SFE are found in regions with the
highest dense gas fractions.

\end{abstract}


\section{Introduction}

Free-free emission from \HII\ regions at radio wavelengths is a good
tracer of young and heavily embedded regions of massive star
formation such as luminous starbursts and nuclear star forming
regions, where visual extinctions can be tens
\citep{1990ApJ...349...57H} to hundreds of magnitudes, affecting
even near and mid-infrared observations. Extragalactic continuum
emission is complicated by the presence of non-thermal synchrotron
emission within the beam, particularly at wavelengths longer than
6~cm \citep{1992ARA&A..30..575C}. While it is possible in theory to
do a spectral and spatial separation of free-free and synchrotron
emission \citep{1983ApJ...268L..79T,1994ApJ...421..122T,2002MNRAS.334..912M} with
multi-wavelength observations, the free-free emission can be most
easily measured at millimeter wavelengths, where radio synchrotron
has declined and dust has not yet begun to take over. The enhanced
high frequency capabilities of the Very Large Array (VLA) --- improved K and
Q band receivers, fast-switching calibration techniques --- allow us
to detect and resolve the free-free emission from large \HII\ regions in nearby galaxies.

We report VLA observations of the 7~mm continuum emission of the
starburst galaxy M82. An interacting dwarf with a bar and gas ring
in the central kpc \citep{1991ApJ...369..135T, 1994ApJ...420..159L,
1995ApJ...439..163A, 1995ApJ...445L..99S}, M82 is a prodigious
former of stars, at a rate of $\sim 10~M_{\sun}~yr^{-1}$
\citep{1988ARA&A..26..343T}. The burst of star formation may have
been triggered by the interaction with M81
\citep{1993ApJ...411L..17Y}, or by infalling molecular gas driven by
the stellar bar \citep{2002A&A...383...56G}. Its current active star
formation can be traced back to 50~Myr ago
\citep{2001A&G....42d..12D}. The strong star formation activity is
probably responsible for the hot gas outflows found in optical, mm,
and X-ray studies \citep{1978ApJ...221...62O, 1987PASJ...39..685N,
1990A&A...228..331L, 2000AAS...196.5002M, 2000ApJ...545L.107M,
2003MNRAS.343L..47S}.

Our goal is to use the 7~mm maps to isolate compact \HII\ regions in
M82, and to determine their free-free flux density, from which we can
infer Lyman continuum rates, or $N_{Lyc}$. With the subarcsecond
resolution afforded by the VLA we can determine locations and
sizes of the bright \HII\ regions on scales of a few pc. Extinction
is high and patchy in M82 \citep{1979ApJ...227...64S}, estimated to
be $A_V \sim 50$ \citep{2001ApJ...552..544F} on large scales, and as
much as several hundred mag behind the molecular clouds
\citep{2005ApJ...635.1062K}, thus radio free-free emission is a
particularly valuable tracer of ionized gas in this dusty starburst
galaxy.

Spectral energy distributions (SEDs) of the compact radio sources at
longer, cm wavelengths \citep{1997MNRAS.291..517W, 1998ApJ...502..218A, 1999PhDT........10A,
2002MNRAS.334..912M} in M82 have shown them  to be mostly non-thermal synchrotron sources,
probably supernova remnants (SNR) but possibly synchrotron-emitting
wind driven bubbles \citep{2007ApJ...659..347S}. 
The structures and the expansion properties of these non-thermal sources have been revealed by Very Long Baseline Interferometry (VLBI) and Multi-Element Radio Linked Interferometry (MERLIN) with 3 -- 35 mas resolutions and multi-epoch monitoring at 18 and 6~cm \citep{1999MNRAS.307..761P,2001MNRAS.322..100M,2006MNRAS.369.1221B,2008MNRAS.391.1384F}.
In addition to the radio continuum work, 
\citet{2004ApJ...616..783R} studied the
H92$\alpha$ and H52$\alpha$ radio recombination lines in M82,
including the 7~mm (42.952~GHz) continuum, in $0.6\arcsec$ or $\sim$
10~pc resolution. 

In this paper, we have 7~mm images at two
resolutions: high resolution, $0.2\arcsec $, for compact structures
on scales of 3 pc, and low resolution, $1.5\arcsec$, for structure on
scales of $\gtrsim$ 25 pc. The map with $0.2\arcsec$ synthesized beam 
is the highest resolution map of M82 at millimeter wavelengths so far. 
Lower resolution maps are important for
judging the amount of missing, undersampled emission in higher
resolution VLA images, which act as spatial high-pass filters.

The distance to M82 was recently determined to be 3.6$\pm$0.3 Mpc
from the optical study of Cepheids in M81 using \HST\
\citep{1994ApJ...427..628F}, assuming the same distance to both
galaxies. Observations of the tip of the red giant branch in M82
itself suggests 3.9 $\pm$0.3 (sys) $\pm$0.3 (random) Mpc
\citep{1999ApJ...526..599S}. For consistency with previous work, we
adopt the 3.6 Mpc distance, with 1\arcsec $=$17 pc. At this distance
a 1 mJy thermal source at 7~mm represents an \HII\ region ionized by
the equivalent of $\sim$ 150 O7 stars.

\section{Observations}

The radio data were acquired with the NRAO VLA{\footnote{The National Radio Astronomy Observatory is a
facility of the \textit{National Science Foundation} operated under
cooperative agreement by Associated Universities, Inc.}} using A-
and B- configurations at 43~GHz (NRAO program ID: AT303) and
D-configuration at 45~GHz (AM839, PI: Meier, D.~S.). Weather during
the A-array observations in October 2004 was not good enough for
useful 7~mm work.

The ``high resolution'' (sub-arcsecond) maps we present in this
paper were made in the B-configuration on 22 April 2005 in continuum
observation mode (effective bandwidth $\sim$172~MHz). The
calibration cycle time for fast-switching was set to 100--120
seconds. 3C147 was used for calibration of the absolute flux 
scale and 0958+655 was the
phase calibrator. Uncertainty in the absolute flux density scale is
estimated to be $<$5\%, based on VLA estimates of the bootstrap
accuracy at 7~mm. For most of our sources, statistical uncertainty
due to noise dominates the uncertainty in flux density scale. Total on source time
is 1.3 hours. The (\textit{u,v}) sampling covers 30--1610
$k\lambda$, resulting in a $\sim$0.2\arcsec\ beam, the size of which
varies slightly with weighting. For our B-array map, the estimated $\theta_{max}$, 
the largest angular scale of structures that our observation is sensitive to, is about $3.5\arcsec$.

Lower resolution 7~mm continuum observations (henceforth ``low
resolution maps'') were made with the VLA D-array at 45.47 GHz on
November 27 2005 in spectral line mode, as part of a spectral line
program (AM839). The setup was two 25 MHz spectrometer tunings per IF
set side-by-side. The uv dataset was generated by fitting the uv
spectrometer data minus the edge channel with a first order
polynomial using {\sc uvlsf} 
task. The net bandwidth is $\sim$
31~MHz. Because IF~2 has contamination from a recombination line,
only IF~1 data is included. The observations used fast-switching
with 180 seconds cycle time. 3C147 and 3C286 were used as flux density
calibrators, and 0958+655 was the phase calibrator. Total
integration time on source is 2.2 hours. The sampled (\textit{u,v})
plane covers 5-157 $k\lambda$, resulting in a $\sim$1.5\arcsec\
beam. We estimate the $\theta_{max}$ is $\sim$ $20\arcsec$ in our D-array map.

The maps were calibrated and reduced in {\sc aips}
following high frequency reduction procedures, and
were deconvolved with {\sc clean}.
The noise level in
the maps is $\rm \sim 0.1~mJy\,bm^{-1}$ for the high resolution
(B-configuration) maps and $\rm \sim 0.5~mJy\,bm^{-1}$ for the lower
resolution (D-configuration) maps, but varies slightly depending on
weighting. Flux measurements and positions were obtained with
{\sc imfit} task in {\sc aips}
unless otherwise noted. 
The {\sc aips} task {\sc pbcor} was used to correct the flux density scale of these data for the primary beam efficiency of the instrument.

\section{Overview: Compact and Diffuse Emission Sources at 7~mm in M82}\label{section:overview}

\begin{figure}
%\figurenum{1}
%\epsscale{0.9}
\plotone{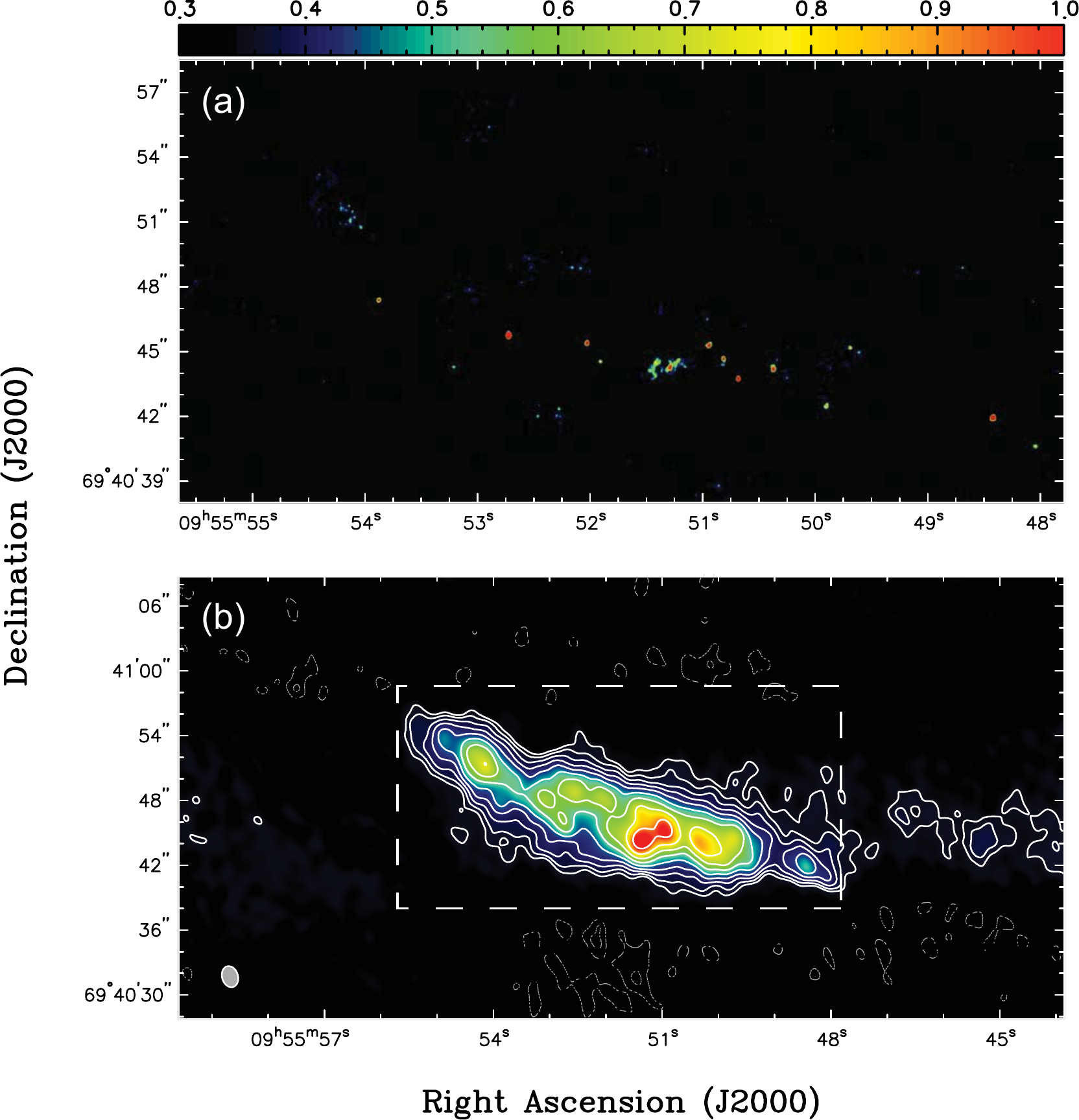} \caption{M82 radio continuum
maps at 7 mm. \textit{(a)}\ Naturally-weighted map from the VLA B-configuration data. 
The color scheme is indicated by the color wedge on the top, with flux density values in mJy/beam.
The synthesised beam is $0.19\arcsec \times 0.15\arcsec$, PA= $-10\degr$. This map is sensitive to structures up to $\theta_{max} \sim 3.5\arcsec$. The peak flux is 5.2 mJy/bm. \textit{(b)}\ Robustly-weighted map made from the D-configuration data.
Synthesized beam is  $1.99\arcsec \times 1.47\arcsec$, PA = $17\degr$ beam. The $\theta_{max}$ is $\sim$ 20$\arcsec$. Contours are
half-integral powers of $2 \times 1.12$ mJy/bm. Dashed box shows the region displayed in (a).}\label{figure:NA_maps}
\end{figure}

The high-resolution B-array map (Fig. \ref{figure:NA_maps}a) covers
the inner 700~pc $\times$ 350~pc ($40\arcsec \times 20\arcsec$) of M82
with a $0.19\arcsec \times 0.15\arcsec$ ($\sim 3$~pc) beam.
Structures larger than $3.5\arcsec$ are suppressed in this map. We
identify 14 compact sources with flux densities $>$ 0.5 mJy
($T_b$=12~K; 5 $\sigma$), distributed in a $\sim$ $40\arcsec \times
10\arcsec$ east-west swath with position angle along the optical major
axis \citep{1972ApJ...173L..47K}. The compact sources are high
brightness, dense, and luminous \HII\ regions and young supernova
remnants (SNR). For the \HII\ regions, our 5$\sigma$ detection limit
corresponds to $N_{Lyc}\sim 0.6 \times 10^{51}~s^{-1}$ (assuming optically thin free-free emission), corresponding to about 80
O stars, or a cluster such as R136. The compact optical \HII\ region M82-A1
studied by \citet{2006MNRAS.370..513S} falls just below our detection
limit.

\begin{figure}
%\figurenum{2}
%\epsscale{0.75}
\plotone{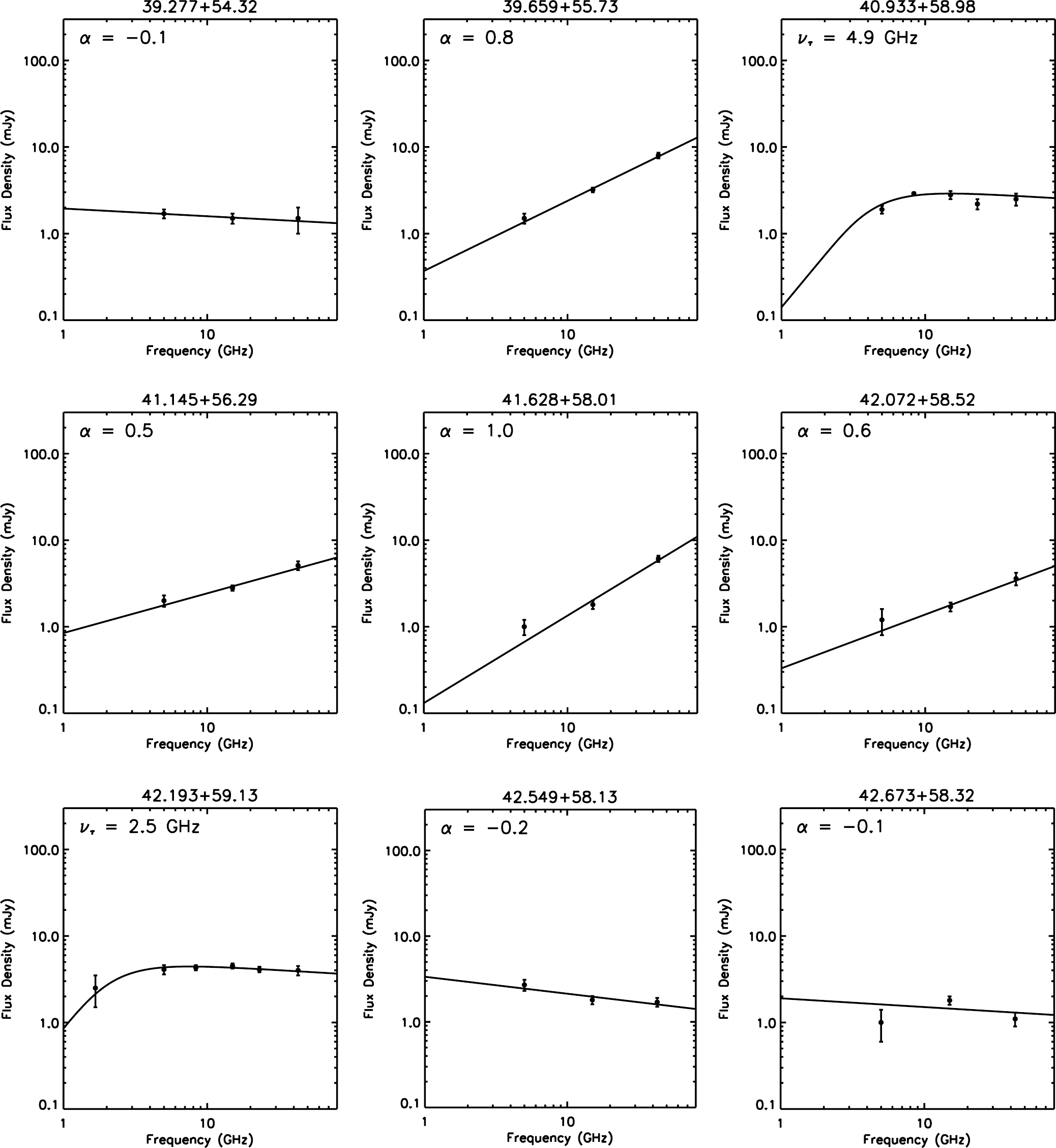} \caption{SED of 9 thermal
sources found in 7~mm high resolution ($0.2\arcsec$) map. Line
represents the best fit single power law SED except in the cases of
40.933+58.98 and 42.193+59.13 where the line indicates free-free
self-absorption.}\label{figure:SED_HII}
\end{figure}

\begin{figure}
%\figurenum{3}
%\epsscale{0.75}
\plotone{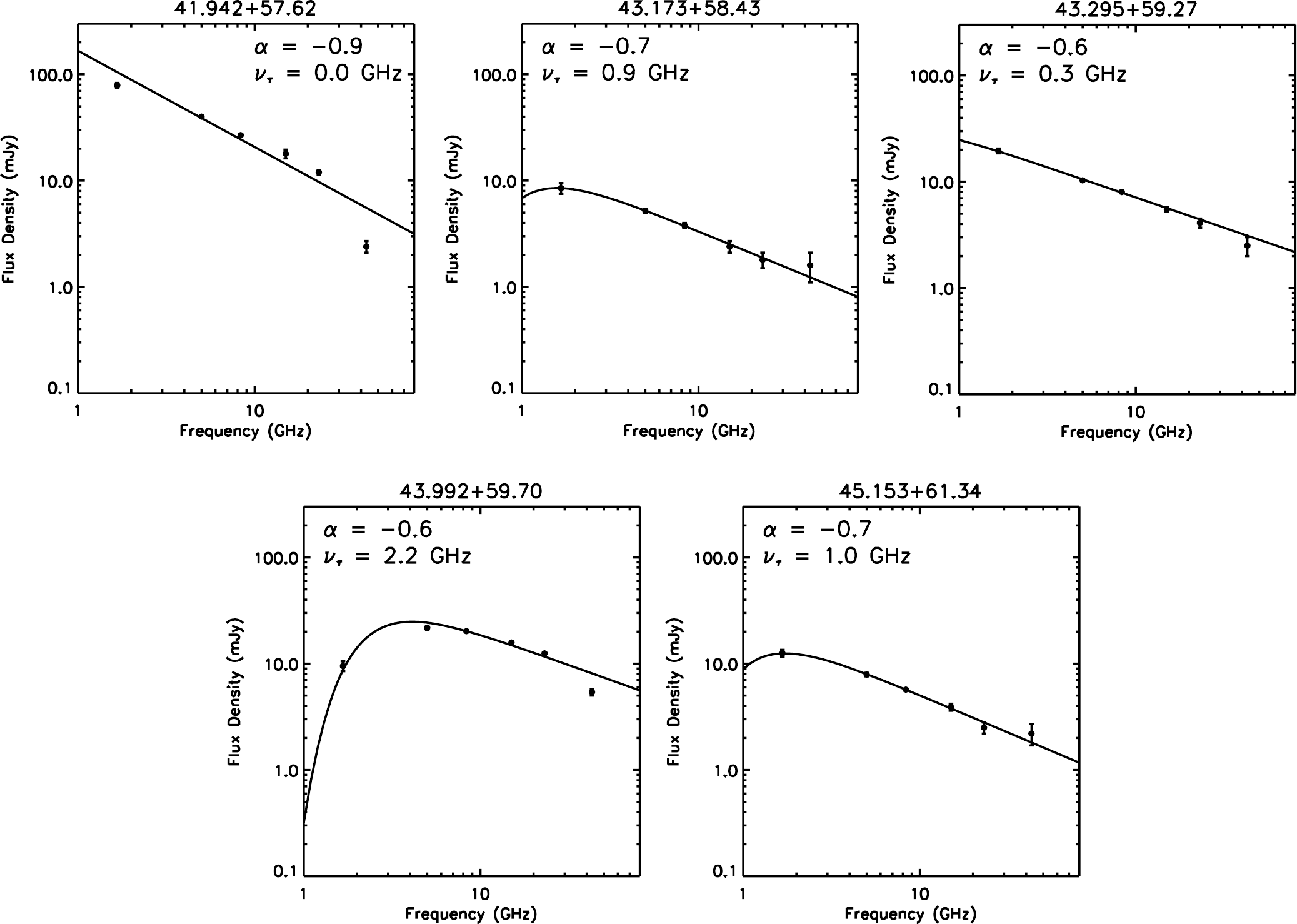} \caption{SED of 5 non-thermal sources
found in 7~mm continuum high resolution ($0.2\arcsec$) map. Line
represents power-law synchrotron radiation with free-free absorption.}\label{figure:SED_SNR}
\end{figure}

Properties of the compact sources in natural weighted map are given in Table
\ref{table:flux_compact}. We adopt the source designations of
\citet{1985ApJ...291..693K}, using the B1950 coordinate of each
source relative to $9^{h}51^{m}+69\degr54\arcmin$, to facilitate
comparisons with earlier studies, but the J2000 coordinates are
given also. In Table \ref{table:flux_compact_200mas}, we have
collected the radio continuum flux densities of all 14 compact sources from
the literature \citep[18 -- 1.3~cm][]{1984MNRAS.211..783U,1985ApJ...291..693K, 1998ApJ...502..218A, 1999PhDT........10A, 2002MNRAS.334..912M} and this work. The flux density measurements at different wavelengths are done at similar resolutions. 
The beam sizes of maps at 18~cm to 1.3~cm range from $0.2\arcsec$ to $0.3\arcsec$. 
The 7~mm measurements quoted in Table \ref{table:flux_compact_200mas} (as well as data shown in Figure \ref{figure:SED_HII}, \ref{figure:SED_SNR}) are made on a tapered map of $0.2\arcsec$ resolution in order to match beam sizes at other wavelengths from the literature. The quoted numbers in Table \ref{table:flux_compact_200mas} are total integrated flux densities of compact sources, and the effect of different beam is less than 30\%. 
In Figure \ref{figure:SED_HII}, \ref{figure:SED_SNR}, and Table
\ref{table:flux_compact_200mas}, we show the SEDs of all 14 compact
sources at wavelength 18~cm to 7~mm. The details of model fitting
are included in Table \ref{table:compact_em}. Our discussion of properties of these compact sources, including their SEDs, can be found in \S\ref{section:compact_thermal} and \S\ref{section:nonthermal_compact_rc}.

The high-resolution B-array map also reveals four regions of
comparatively extended emission ($EM = \int n_{e}^{2}dl$). The bright, high frequency
free-free emission detected here arises from gas with very high
emission measure, so it is surprising to find such extended sources.
They are 1--2\arcsec\ across, or 17--35 pc. They are similar in
extent to the large \HII\ region associated with NGC~3603 in our Galaxy, 
but with a much higher brightness and EM. We discuss the
significance of these high brightness, extended regions in
\S\ref{section:extended_EM}.

The low resolution D-array map is shown in Figure
\ref{figure:NA_maps}b. This map, with a beam of 1\farcs99
$\times$1\farcs47 p.a. 17\degr, is sensitive to lower brightness emission,
$T_b\sim 0.1$~K rms ($\sigma\sim$0.5 mJy/bm) and also more extended
emission structures (up to 60\arcsec, covering the entire field
shown) than the high resolution map. The low resolution image shows
four major peaks with intensity 12~mJy/bm or higher.
The integrated flux density in the naturally weighted D-array map
is $1.2\pm0.1$ Jy. The total flux reported for M82 from
\textit{WMAP} at 41 GHz flux density is $1.3\pm0.2$ Jy
\citep{2003ApJS..148...97B}. Thus our low resolution D array map
recovers, within the uncertainties, all the 7~mm flux density. The diffuse emission structure is 
elongated in the same direction and over the same extent as the
swath of compact sources, and agrees with the orientation of the
stellar elliptical component seen in the deep near infrared images
\citep{2005ApJ...628L..33M}.

We have fitted 17 peaks in the low resolution map with Gaussians, and
the corresponding flux densities, locations, and the sizes of the peaks are
listed in Table \ref{table:peaks_Darray}. All of these peaks are found in the 
longer wavelength  VLA images of 
\citet{1999PhDT........10A} except  peaks W4a  and E1a, which are newly identified in this work. The peaks sum to $\sim$
0.65~Jy at 7~mm. If this flux density associated with low resolution peaks
is all optically thin free-free emission, and $T_{e} =
10,000~K$, the inferred Lyman continuum photon rate ($N_{Lyc}$) is
$\sim 8.4 \times 10^{53}~sec^{-1}$. (For $T_{e}$ of 5,000, $N_{Lyc}$  
would increase by 30 -- 40\%.) This value is in good agreement with
that estimated from the free-free continuum at longer wavelengths by
\citet[][$9.1 \times 10^{53}~\rm sec^{-1}$, from 1.3 -- 20~cm
multi-wavelength spectral decomposition
modeling]{1999PhDT........10A}, and at shorter wavelengths by
\cite{1991ApJ...366..422C} and \cite{2005ApJ...618..712M}, who find
$6.9 \times 10^{53}~\rm sec^{-1}$ and $7.9 \times 10^{53}~\rm
sec^{-1}$ at 3~mm and 2.6~mm, respectively (these numbers are
interferometric and therefore are lower limits). It also agrees with
that suggested from recombination line studies
\citep{1985ApJ...294..546S} and the extinction-corrected
mid-infrared fluxes of \citet{2001ApJ...552..544F}. The good agreement in $N_{Lyc}$ between our
analysis and previous studies indicates that the peaks we see in the
low resolution map of Figure \ref{figure:NA_maps}b dominate the total free-free emission.

Non-thermal emission at 43 GHz is weak, but contributes a
non-negligible fraction of the integrated flux density in M82. The estimate of the non-thermal
component is 35\% of the total at 7~mm, extrapolating from VLA cm-wave flux densities of the same region
\citep{1999PhDT........10A}. This would give a thermal flux density of $\sim
0.8~Jy$, 15 -- 20\% higher than 0.65~Jy we obtain from summing the peaks in the D-array map of 
Figure \ref{figure:NA_maps}b. We attribute the discrepancy to thermal emission from some \HII\  
regions distributed on too large a spatial scale to be seen in peaks in these
images.

\section{Star Formation Tracer: 7~mm Continuum Emission in M82}

\subsection{The Compact Thermal Radio Sources}\label{section:compact_thermal}

The 14 compact sources of the high resolution map (Fig.\
\ref{figure:NA_maps}a) contribute $\lesssim 10\%$ of the total
7~mm flux density in the low resolution D-array map (Fig.\
\ref{figure:NA_maps}b). These compact sources in the
7~mm high resolution B-array map also appear in lower frequency
radio continuum maps 
\citep{1994MNRAS.266..455M, 1994ApJ...424..114H, 1997MNRAS.291..517W, 
1998ApJ...502..218A, 1999PhDT........10A, 2001MNRAS.322..100M, 
2008MNRAS.391.1384F}. We have computed radio SEDs for the
14 sources, which are shown in Figures \ref{figure:SED_HII} and
\ref{figure:SED_SNR}, with details of the SED modeling given in Table
\ref{table:compact_em}. We identify nine regions with spectral index
$\alpha \gtrsim -0.2$ ($S \propto \nu^{\alpha}$) as \HII\ regions,
and the others as SNR. The flux densities at 43 GHz and the spectral
indices are consistent with previous SEDs and source characterizations 
\citep{1998ApJ...502..218A,2002MNRAS.334..912M} except for the two
non-thermal sources 43.992+59.70 and 41.942+57.62, which are
discussed in \S\ref{section:AGN} and
\S\ref{section:individual_nonthermal}, respectively. 

Of the nine \HII\ regions in the high resolution map, two
have flux densities at more than five frequencies. Fitting of the
spectra of these two sources suggests that they 
have free-free turnover frequencies (where $\tau \sim 1$)
$\nu_t \sim $2.5--5~GHz. To have $\nu_t \sim$ a few GHz requires
emission measures of $2 - 9 \times 10^{7}$~pc~cm$^{-6}$. 

The other sources have flux densities at only three frequencies, and a single
power law represents their observed flux densities fairly well (Fig.\
\ref{figure:SED_HII}). 
Some of the deviations may be due to small amounts of extended emission in the sources
and the slightly (15-20\%) differing beams of the flux densities from the literature. 
Four sources have rising SED ($\alpha > 0$).
Rising spectrum radio sources have been found in many extragalactic
systems, and are thought to be giant compact \HII\ regions with high
electron density \citep{1998AJ....116.1212T}. These are much larger than Galactic compact \HII\ regions because the super star clusters (SSCs), the clusters exciting them, are much larger both in excitation and in physical size. For the sources here,
the emission appears to remain partially optically thick at frequencies up to
43~GHz, so we expect to see brightness temperatures of $10^2 - 10^3$~K. Their observed brightness temperatures are lower than this, and imply EMs $\gtrsim 3.5 \times
10^{7}~pc~cm^{-6}$ (derived from peak flux density), consistent with turnover frequencies of
3~GHz, an order of magnitude lower than observed. The low brightness temperatures indicate small filling factors, and signal the
presence of very compact sources of significantly less than 3 pc extent, which we cannot resolve.  

Considering the radio continuum luminosity, brightness of the
emission, and the turnover frequency, these appear to be young \HII\
regions, and we infer that the star clusters exciting these nebulae
are also relatively young. The total flux density of just the detected compact
\HII\ regions comes to $\sim$ 38 mJy, which is $\sim$ 6\% of the total
thermal flux density at 7~mm (at the same sensitivity), in other words, $\sim$ 6\% of the ionizing
flux density. But the high resolution 7~mm image fails to detect all of the
compact \HII\ regions in M82, even fairly optically luminous ones such as M82-A1
\citep{2001MNRAS.326.1027S}, so we believe that there are more compact
regions below our current detection limit. In this case the 
\HII\ region phase in M82 must last at least $\sim$ 6\% of the $\sim$10 Myr lifetime of O stars,
and possibly longer, to a Myr or more.  To restrain the \HII\ regions from expanding requires
external or internal confinements, perhaps external high pressures \citep{1995RMxAA..31...39D, 2006MNRAS.370..513S}, or in
the larger clusters, potentially slowing by gravity
\citep{2003Natur.423..621T}.

\subsection{Lack of Optical/Near-IR Continuum Counterparts}

\begin{figure}
%\figurenum{4}
\epsscale{1.0} \plotone{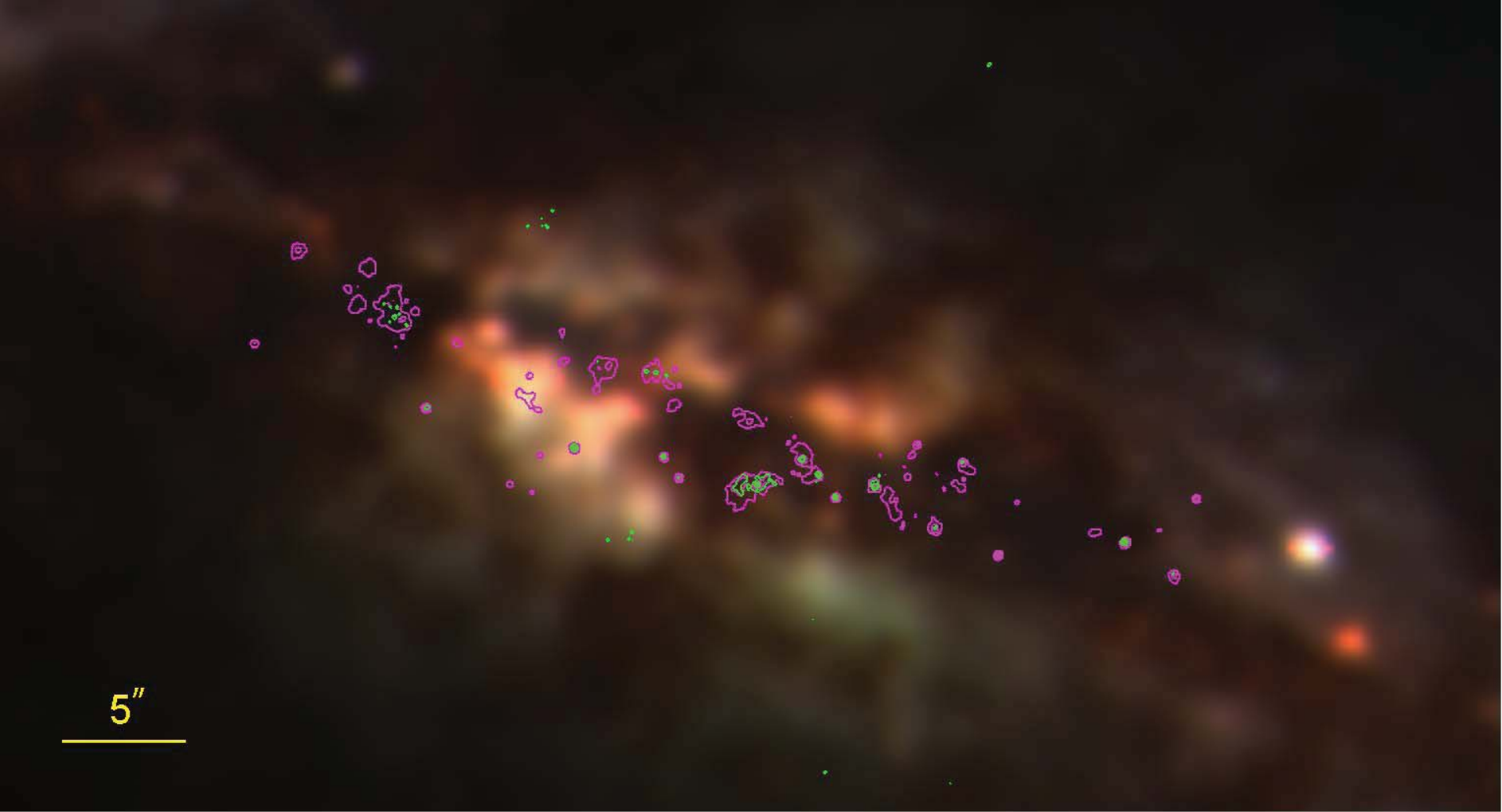} \caption{Compact radio
sources overlaid on false color VRI image from \Spitzer\ SINGS
project. The magenta contours at integers $\times$ 0.5 mJy/beam show 2~cm emission reproduced from VLA
archive data (AM671; \citealt{2002MNRAS.334..912M}) at resolution
$\sim 0.2\arcsec$. The green contours at integers $\times$ 0.8 mJy/beam present the 7~mm sources shown
in Figure \ref{figure:NA_maps}a.}\label{figure:dust_lanes}
\end{figure}

The compact sources are shown in Figure \ref{figure:dust_lanes}
overlaid on a color (VRI) image from the \Spitzer\ Legacy project
SINGS \citep{2003PASP..115..928K}. Nearly all of the compact sources
are found in dust lanes. The only exception is source 43.922+59.70,
which is located near the dynamical center,
and a potential AGN candidate ($\S$6.) 

We have attempted to align the compact radio sources identified in
this and previous work (Allen 1999, McDonald et al 2002, and
references therein) with the published \HST\ near-IR continuum images
of \cite{2003AJ....125.1210A} and \cite{2003ApJ...596..240M}, the
catalogued optical clusters \citep{2005ApJ...619..270M}, and the
archived \HST\ optical measurements. There is no consistent alignment
of the radio and other sources within the \HST\ pointing
uncertainty. The \HII\ regions we see are presumably
associated with young SSCs; that the clusters are not
seen in the optical/Near-IR continuum argues that the extinction must be high
($A_{K} > 6$). The molecular clouds alone can contribute $A_{V}$ of $10^{2}$--$10^{3}$ if they happen to be along the line of sight to the \HII\ regions. 

If the average extinction at arcsecond  
resolution is $A_{V} \sim 14$ \citep[from H-K color by][assuming 
$A_{V}/A_{K}=10$]{1993ApJ...412..111M}, the
line-of-sight extinctions implied in the lack of near-IR sources to
these young clusters must be 4 times higher. This is consistent with
the high average ($A_{K}\gtrsim 1$) and patchy extinction derived on
large (11\arcsec , or $\sim$ 200 pc) scales toward the \HII\ regions
in M82 from Brackett line observations by 
\citet[][$A_{V} =25$]{1977ApJ...217L.121W}, 
\citet[][$A_{V} =14$]{1979ApJ...227...64S},
\citet[][$A_{V} =26$]{1980ApJ...238...24R},
and \citet[][$A_{V} \sim
50$]{2001ApJ...552..544F}, much of which appears to be internal to the
\HII\ regions themselves. 

In the $Pa\,{\alpha}$ image of \cite{2003AJ....125.1210A},
we identify 3 (39.227+54.32, 40.933+58.98, and 42.072+58.52) out of our
9 compact thermal sources, and one non-thermal source (43.992+59.70) as associated
with compact $Pa\,{\alpha}$ structures. If the $Pa\,{\alpha}$ emission is associated with the corresponding radio
\HII\ regions, we estimate, based on the inferred Lyman continuum
flux densities, that the extinction toward these IR-visible nebulae is $A_{K}
\gtrsim 9$. None of these three sources are apparent in the \HST\ near-IR
continuum images, and it agrees with the high extinction suggested
in the $Pa\,{\alpha}$ image.  

\subsection{High EM Extended Structures at 7~mm: Sheets of Hot Gas}
\label{section:extended_EM}

\begin{figure}
%\figurenum{5}
\epsscale{0.6} \plotone{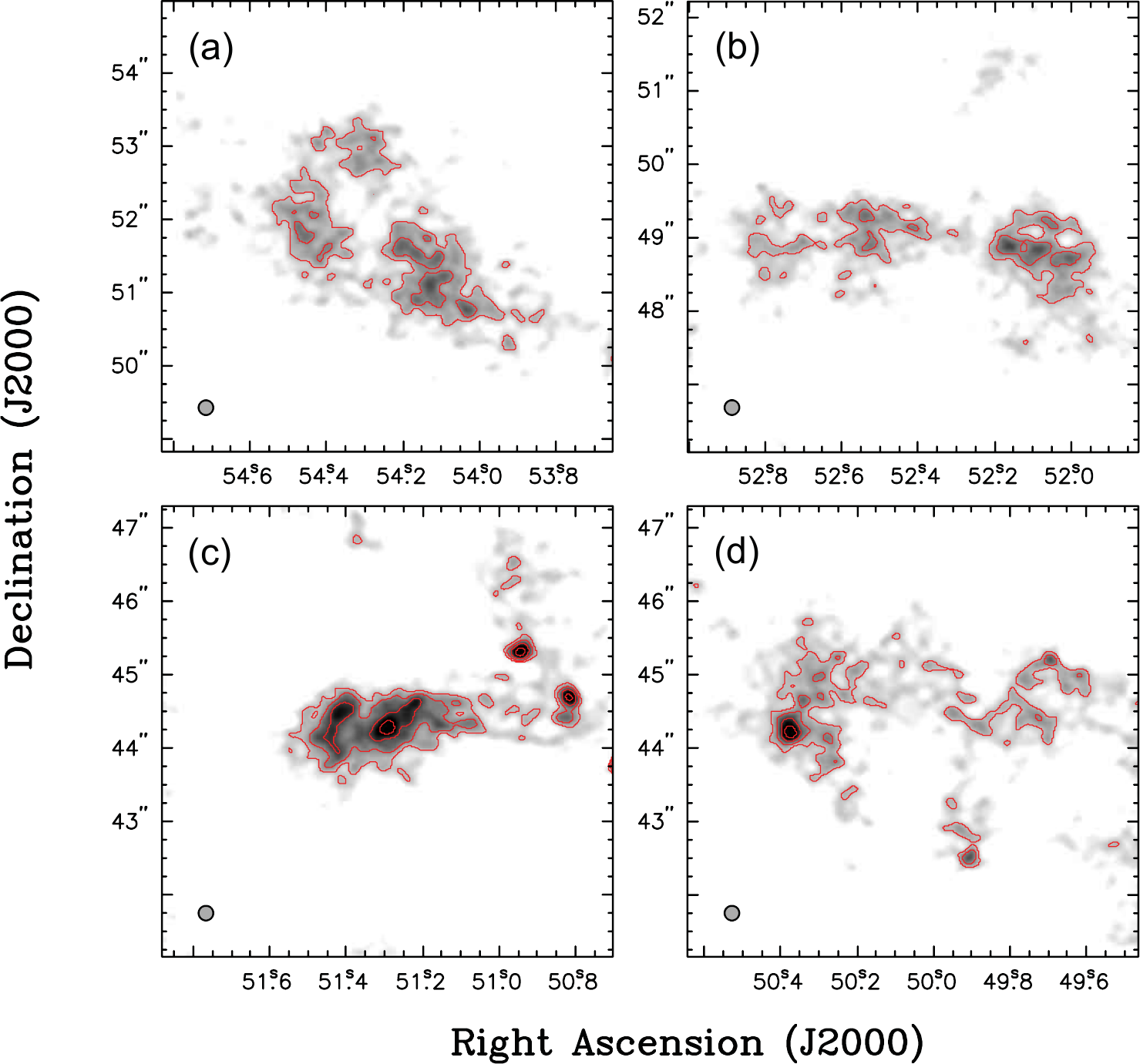} \caption{7~mm
continuum emission from extended sources in M82 at $09^{h}
55^{m};+69\degr 40\arcmin$. The contours are $2^{\frac{n}{2}} \times
0.6 mJy/bm$ in integer $n$. Maps are generated with (\textit{u,v})
restriction of 30 -- 1600~$k\lambda$ in $0.2\arcsec \times
0.2\arcsec$ beam.}\label{figure:extended_B-array}
\end{figure}

Besides the compact dense sources in Table \ref{table:compact_em},
we detect extended regions in M82 with relatively high EM. In Figure \ref{figure:extended_B-array}
we show B-array maps of free-free emission regions with integrated
flux density of 50 -- 70~mJy in 5\arcsec -- 9\arcsec\ area.
These regions have $EM = 1.2 - 2.0 \times
10^{7}~pc~cm^{-6}$. This agrees with the D-array measurements shown
in Table \ref{table:peaks_Darray}.

High EM can be produced either by high rms electron density or by a long sightline ($dl$) of ionized gas. If high electron density is responsible, then to reconcile the high EM with the observed flux ($N_{Lyc}$) requires that the line of sight length be much smaller than the observed transverse distance. For example, to maintain the ionization of a spherical volume of radius 15~pc, the observed transverse diameter of source in Fig \ref{figure:extended_B-array}a, and rms electron density $10^3~cm^{-3}$ ($\gtrsim 5 \times 10^{3}~cm^{-3}$ was inferred by \citealt{2004ApJ...616..783R} in a 2\arcsec\ resolution) requires a Lyman continuum rate of $N_{Lyc} \sim 2 \times 10^{53}~s^{-1}$. The observed flux of this structure indicates $N_{Lyc} \sim 10^{52}~s^{-1}$. This implies that the actual emitting volume is smaller, i.e., the emitting region is sheetlike. There are a number of such patches evident in Figure \ref{figure:extended_B-array}. While these large regions of high electron density are dynamically stable in the high pressure $\rm P/k \gtrsim 10^7~cm^{-3}~K$ environment of the M82 starburst, the excitation requirements are inconsistent with other star formation indicators, with the possible exception of the ``yin-yang'' shaped source (Fig.\ \ref{figure:extended_B-array}c) located at the peak of the low resolution map (Fig.\ \ref{figure:NA_maps}). These structures could be the ionized surfaces of  chimneys \citep{1999MNRAS.309..395W} or superbubbles \citep{1998A&A...339..737N,1999A&A...345L..23W,1999MNRAS.309..395W,2000ApJ...545L.107M,2005ApJ...618..712M}. The coincidence of their locations with dust lane and CO gas distribution implies that they are associated with in the molecular clouds, which would be consistent with an origin in the outer, ionized layers of large molecular clouds.

Despite its location near the edge of the superbubble, the ``yin-yang'' source is an exception in that it does appear to have sufficient flux to support the ionization, and there is a bright mid-IR source here too \citep{1995ApJ...439..163A,2004ApJ...603...82L}. However this source has an unusual structure that is coherent and bubble-like on a scale of $1.5\arcsec$ = 26~pc. If this were a bubble, an energy of $10^{50.5}$~ergs would be required to clear the inner region of gas.

Another explanation for these extended regions is that they represent a collection of smaller \HII\ regions representing distributed star formation at the end of the stellar bar in M82. Regions (b) - (d) in Figure \ref{figure:extended_B-array} are distributed on the 10\arcsec\ ionized ring of M82 \citep{1994ApJ...420..159L,1995ApJ...439..163A}, which may represent the $x_{2}$-orbits of a bar \citep{1991ApJ...369..135T,1994ApJ...420..159L,1995ApJ...439..163A,2002A&A...383...56G}. The region (a) is on the east ionized ridge, and probably near the end of the $x_{2}$-orbits. Star formation is observed to be near bar ends, perhaps facilitated by cloud-cloud collisions.

\subsection{\HII\ Regions, Molecular Gas, and Dust}

\subsubsection{Correlation with Molecular Gas: CO(1-0) and HCN(1-0)
Line Emission}\label{section:Molecular_Gas}

\begin{figure}
%\figurenum{6}
\epsscale{0.75}
\plotone{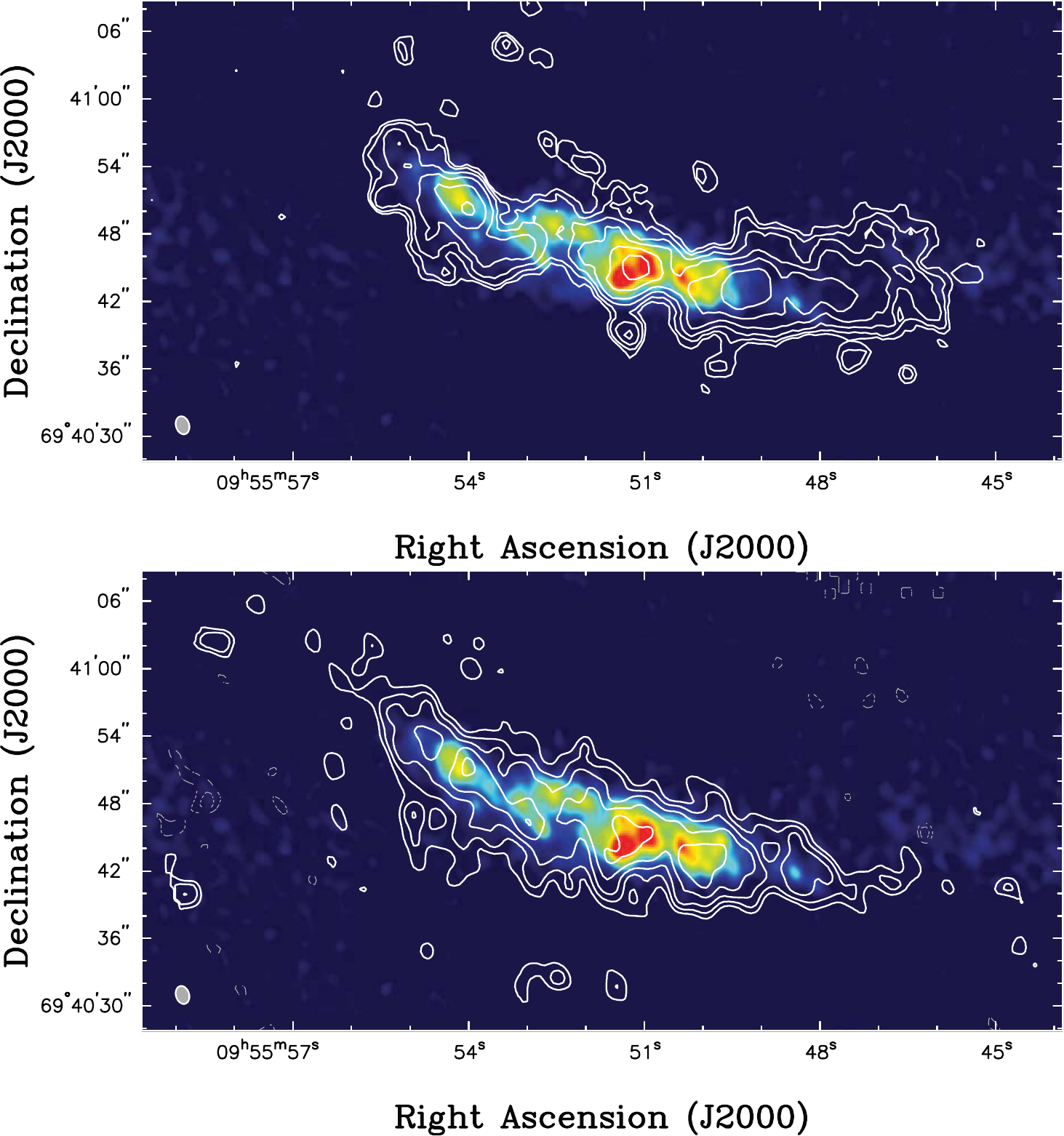} \caption{Molecular line
emission contours overlaid on color images of the D-array 7~mm continuum robust (R=0)
map. The contours on the top are CO(1-0) from Shen \& Lo (1995).
The contours are $2^{\frac{n}{2}} \times 5$ Jy/bm~km/s. On the
bottom is the HCN from Schilke (private communication). The contours
are $2^{\frac{n}{2}} \times 0.8$
Jy/bm~km/s.}\label{figure:Molecular_gas}
\end{figure}

The dense star forming regions traced by high brightness 7~mm
free-free emission are likely to be still close to their natal giant
molecular clouds (GMCs). We therefore expect a high correlation of thermal
free-free emission and molecular line emission, although at some
sufficiently high resolution, they should begin to anti-correlate.
In Figure \ref{figure:Molecular_gas}, we show the 7~mm map with
contours of CO(1-0) line emission from the OVRO map of
\citet{1995ApJ...445L..99S} and an HCN(1-0) map from the Plateau de
Bure Interferometer (Schilke, private communication). The molecular
gas tracer CO(1-0) is more extended in the plane of the galaxy,
while the dense molecular gas tracer HCN(1-0) emission is confined
to the star formation area. It is apparent that the correlation of
the 7~mm continuum emission, largely free-free emission, with 
HCN gas is excellent.

\begin{figure}
%\figurenum{7}
\epsscale{0.55} \plotone{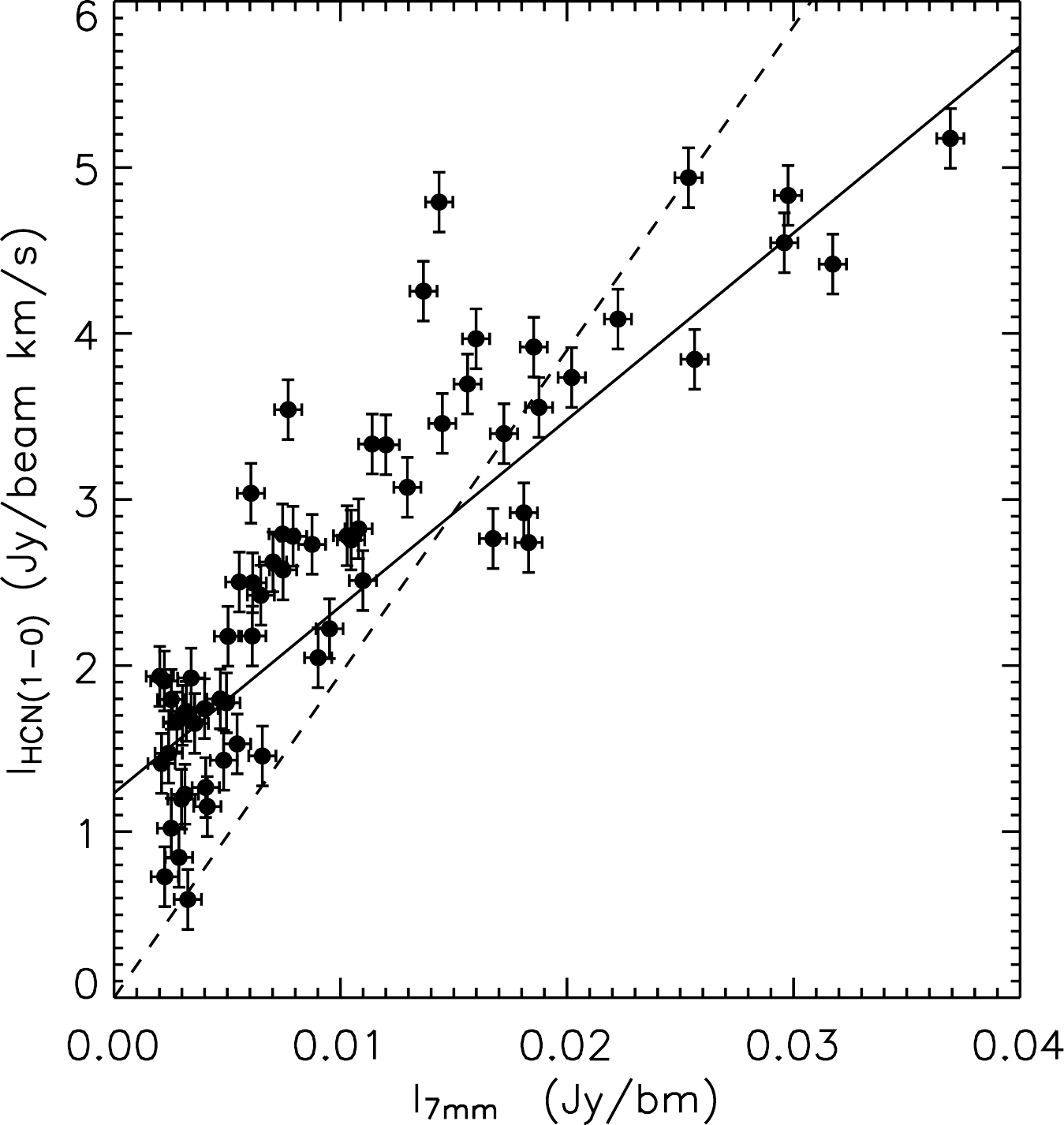} \caption{Intensity
of HCN (J=1-0) line v.s. D-array radio continuum strength at 7~mm. Each
point on the plot represents independent measurements with a
$2.7\arcsec \times 2.7\arcsec$ beam for flux density $> 3~\sigma$. Solid and
Dashed lines represent the linear regression result of $I_{HCN}\sim
112~I_{7mm} + 1.2$ and $I_{HCN}\sim 195~I_{7mm}$ (forced zero
intercept), respectively.}\label{figure:HCN_7mm}
\end{figure}

A quantitative comparison of the molecular line fluxes and the free
free flux confirms this conclusion. Figure \ref{figure:HCN_7mm}
shows the beam-to-beam comparison between the low resolution 7~mm emission and
HCN. The linear correlation between HCN and radio
continuum shown in Figure \ref{figure:HCN_7mm} agrees with the tight
correlation between global HCN emission and star formation tracers
(such as far-IR emission) discovered in local normal spiral and starburst galaxies \citet{2004ApJ...606..271G}, in Galactic molecular dense cores \citep{2005ApJ...635L.173W}, and in high redshift galaxies
\citep{2007ApJ...660L..93G}. Our plot shows that this good
correlation between dense molecular gas and star formation extends
down to scales of tens of pc.  

\begin{figure}
%\figurenum{8}
\epsscale{0.75} \plotone{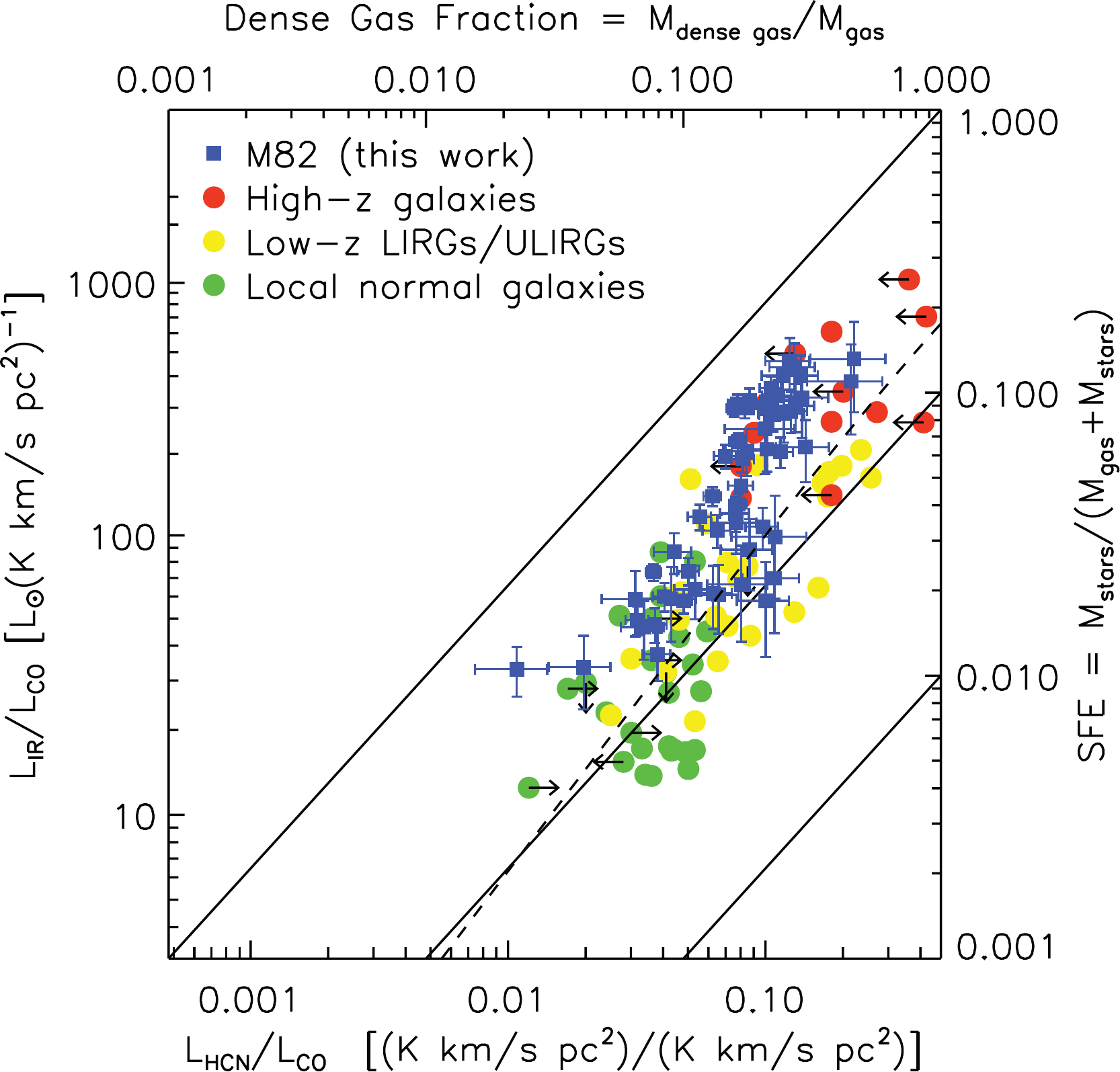} \caption{Correlation between
$L_{IR}/L_{CO}$ and $L_{HCN}/L_{CO}$. Filled blue squares are data
of M82 at $2.7\arcsec$ resolution. Data points of M82 with less than $3~\sigma$ detection are
suppressed. Green ($L_{IR} < 10^{12} L_{\sun}$) and yellow dots
($L_{IR} \geq 10^{12} L_{\sun}$, LIRGs or ULIRGs) are galaxies in
local universe while the red dots are high redshift galaxies
\citep{2007ApJ...660L..93G}. We estimate $L_{IR}$ from the
total 7~mm flux density by assuming the $L_{IR} = L_{OB}$ is dominated by the ZAMS
population that excite the 7~mm \HII\ regions (see text for
details). The dashed line represents the correlation found by \citet{2004ApJ...606..271G}. The axes on the top and right show the corresponding 
dense gas fraction (DGF) and star formation efficiency (SFE). 
The diagonal solid lines from top to bottom represent the ratio between SFE and DGF as 1, 0.1, and 0.01, respectively.}\label{figure:Gao_diagram}
\end{figure}

To more directly compare our results with
\citet{2007ApJ...660L..93G} who determined the correlation in terms
of infrared luminosity, $L_{IR}$ and HCN line flux, we have
translated the 7~mm flux density into $L_{IR}$. We assume an
O-star-dominated stellar population which yields the relation of
$L_{IR} \sim L_{OB} \sim 2 \times 10^{5}~L_{\sun} \times
(\frac{N_{Lyc}}{10^{49}~sec^{-1}})$, following
\cite{2000ApJ...532L.109T} and \cite{2002AJ....124.2516B}, or $L_{IR} \sim
2.9 \times 10^{7}~L_{\sun} \times (\frac{S_{7mm}}{mJy})$ at 3.6~Mpc. This relation could result in underestimates if evolved stellar populations contribute significantly to $L_{IR}$.
Our 7~mm values are plotted with the data of
\citet{2007ApJ...660L..93G} on Figure \ref{figure:Gao_diagram}.
Further discussion of this correlation is in \S\ref{section:SFE}.

\subsubsection{Correlation with \Spitzer\ $8~\micron$ Emission}\label{section:PAH}

The polycyclic aromatic hydrocarbon (PAH) emission features in the mid-infrared
 are expected to trace photo-dissociation regions (PDRs)
generated by the softer UV photons produced by OB stars \citep{1997ARA&A..35..179H}.
Emission in the $8~\mu m$ band of the IRAC camera is dominated by
aromatic feature emission \citep{2006ApJ...642L.127E}, probably
PAHs. The contribution of
starlight to the $8~\mu m$ band is estimated to be minimal, at $\sim
$10\% \citep[based on IRAC 3.6~\micron\
flux;][]{2004ApJS..154..253H}. The resolution of IRAC at this
wavelength, $1.7\arcsec$, is comparable to our low resolution 7 mm
image.

\begin{figure}
%\figurenum{09}
\epsscale{0.9} \plotone{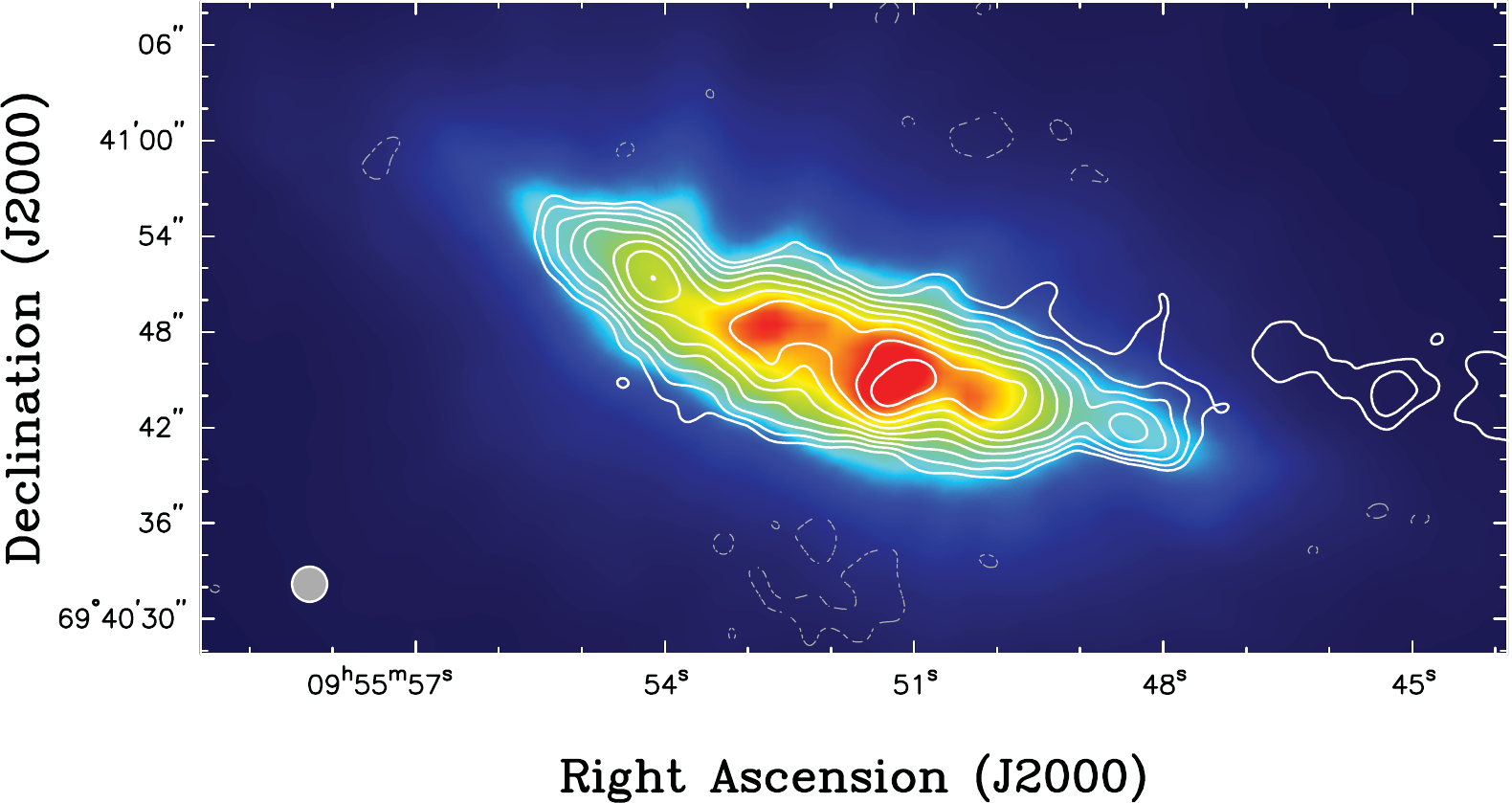} \caption{Naturally
weighted D-array 7~mm continuum contours overlaid on \Spitzer\ IRAC
$8~\micron$ map. The contours are $2^{\frac{n}{2}} \times 1.6$
mJy/bm ($\sim 2~\sigma$). The resolutions of both maps are matched
to 2.2$\arcsec$, the size of PSF in IRAC
$8~\micron$.}\label{figure:PAH}
\end{figure}

M82 was observed at $8~\mu$m by IRAC on \Spitzer\ as part of the
\Spitzer\ Infrared Nearby Galaxy Survey
\citep[SINGS;][]{2003PASP..115..928K}. The 7~mm contours of our low
resolution D-array map are shown overlaid on the IRAC $8~\micron$
map in Figure \ref{figure:PAH}. The 7~mm and $8~\mu m$ emission have
very similar morphologies.

\begin{figure}
%\figurenum{10}
\epsscale{0.75} \plotone{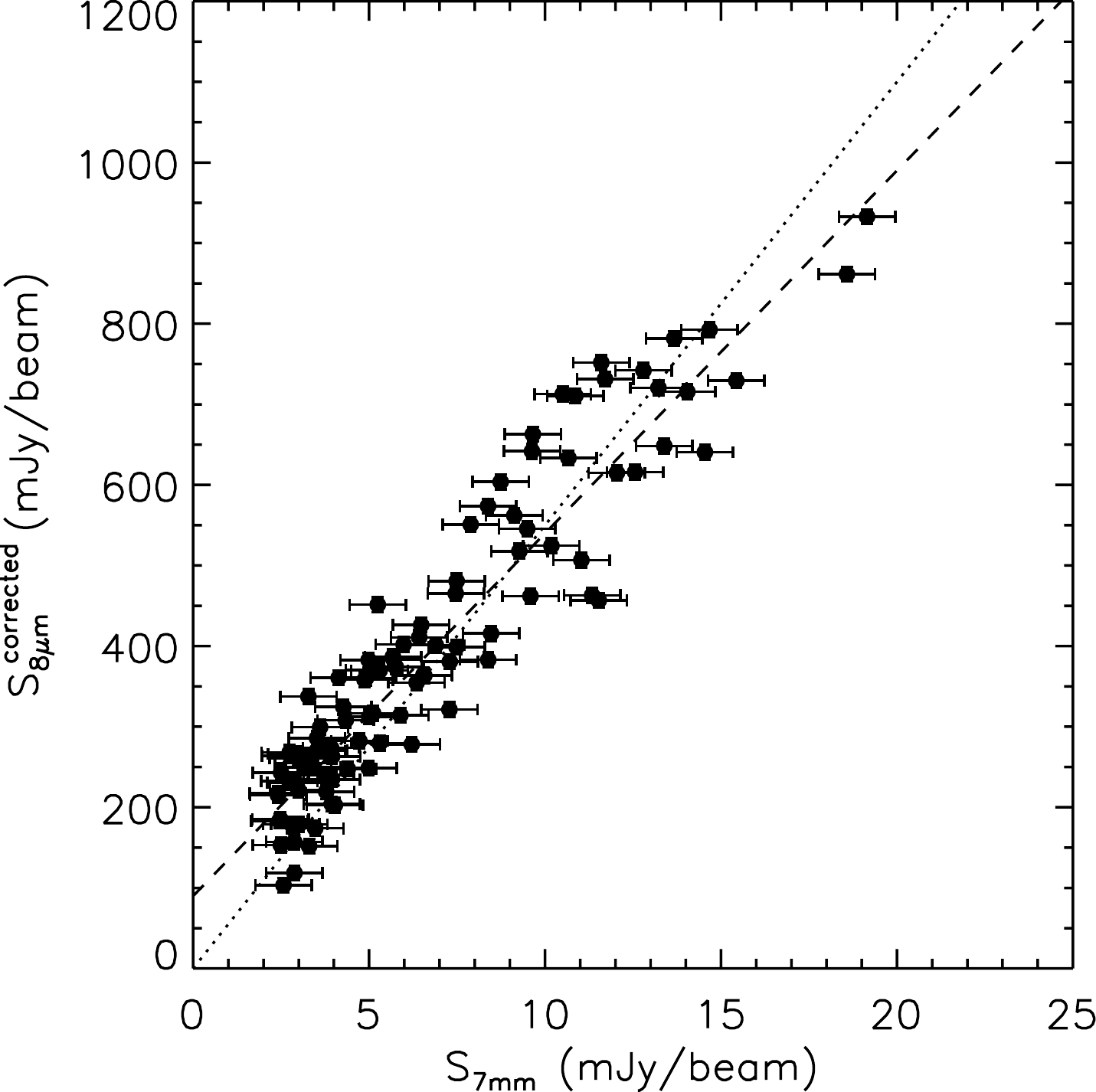} \caption{Correlation between
flux of PAH feature at $8~\micron$ v.s. D-array radio continuum
emission at 7~mm. The plot is made with maps of $2.2\arcsec$
resolution in $2.2\arcsec$ pixels at both wavebands. Each point
represents the measurement at an individual pixel. The PAH flux is
corrected from IRAC measurements. Data under 3~$\sigma$ are
suppressed. The dashed line represent the linear regression result
of $S_{8\micron,corrected}^{IRAC} \simeq 45\times S_{7mm}^{D-array}
+ 90$ in mJy/bm. The dotted line shows the regression of
$S_{8\micron,corrected}^{IRAC} \simeq 55\times S_{7mm}^{D-array}$ in
mJy/bm with a forced zero intercept.}\label{figure:PAH_Free-Free}
\end{figure}

We have compared the corrected flux density in IRAC $8\micron$ and
at VLA D-array 7~mm beam-by-beam. The  IRAC $8\micron$ flux density is
corrected to remove the stellar emission in the IRAC $8\micron$
band, following \citet{2004ApJS..154..253H}. The 7~mm map is
convolved to $2.2\arcsec \times 2.2\arcsec$ to match the PSF of IRAC
channel 4 ($8\micron$). The $I_{8\micron}$ v.s. $I_{7mm}$ plot is
shown in Figure \ref{figure:PAH_Free-Free}.  We limit the data points shown to those with $3~\sigma$ detection at 7~mm. This process reduces the contribution of non-thermal radio emission which is fainter at high frequency and more extended.  
The ratio is linear,
with  $I_{8\micron}/I_{7mm} \approx 45$ and  a $I_{8\micron}$ zero
intercept $\sim 90\pm10~\rm mJy/bm$ in 2.2\arcsec\ beam. The high
correlation of $8~\micron$ and 7~mm emission may mean that they are
excited by the same relatively strong radiation sources on the $\sim
40~pc$ length scale. The $I_{8\micron}$ that appears even at zero
$I_{7mm}$, about 40\% of the total, could be excited by the
relatively older population of B stars that do not produce strong
radio continuum emission.  

Integrating the correlated (above the offset level) $8~\micron$ flux density
gives about 30~Jy. Our beam-to-beam linear correlation between
$I_{8\micron}$ and $I_{7mm}$ suggests that the Lyman continuum rate is
\begin{equation}
N_{Lyc} \sim 2.2 \times
10^{51}~sec^{-1}~(\frac{D}{Mpc})^{2}(\frac{S_{8\micron}^{correlated}}{Jy})
\end{equation}
for $I_{8\micron} > 0.06~\rm mJy/pc^{2}$. 
Other relations between
$S_{8\micron}$ and $N_{lyc}$ have been found from work at different
wavelengths. Deriving a $Pa\alpha$ flux from our $N_{lyc}$ (using
$N_{lyc} = 9.8 \times
10^{63}~sec^{-1}~\frac{S_{Pa\,{\alpha}}}{ergs~sec^{-1}~cm^{-2}}$ from
\citealt{1990ApJ...349...57H} and \citealt{1995MNRAS.272...41S}), the
flux density expected from our 7~mm derived $N_{Lyc}$ using the
\citet{2007ApJ...666..870C} relation for
$S_{8\micron}/S_{Pa\,{\alpha}}$ would give about a factor of 5 more
$S_{8\micron}$ than observed. This may not be outside the large scatter
in the observed relationship, which \citet{2007ApJ...666..870C}
discuss. Another prediction of the $8\micron$ flux density can be made
from the empirical relations $N_{Lyc}/L_{6.2PAH} \sim 10^{46} ~
sec^{-1}/L_{\sun}$ \citep[][based on studies on normal galaxies,
starburst galaxies, and \HII\ regions in the Local Group]{2004ApJ...613..986P} and $L_{6.2~\micron}/L_{8.6~\micron} = 1.5$
\citep{2007ApJ...656..770S}.  This would give $S_{8\micron}$ about a
factor of 10 lower than observed. All relations of the PAH luminosity
to the $N_{lyc}$ are uncertain and have large scatter because the PAH
flux will depend on the geometry of the radiation field, the role of
dust, the presence of B stars, and possibly other factors not yet
considered.  The spatial correlation of the PAH feature and the
ionization in M82 is clear; the quantitative relation is not yet so.

PAH emission may be associated with a B star
population \citep{2004ApJ...613..986P} because the strong UV field
of O stars could destroy PAHs. If we constrain the ionization
sources of the radio continuum emission in \HII\ regions to be B
stars, forcing the initial mass function (IMF) to be deficient in O stars, then to produce
the observed $N_{Lyc}$ would require the total mass of young stars
to increase by 2 orders of magnitude relative to a normal IMF. 
This leads to a star formation rate of few hundred solar mass per year for the past 100 Myr, which
is highly unlikely.

Why do the radio continuum emission peaks correlate with
$8~\micron$ peaks, where the PAH carriers should be destroyed by the
UV? It is possible that dust within the \HII\ regions around the
massive stars absorbs a significant fraction of the UV flux
\citep{2001ApJ...552..544F}. Or the production rate of PAHs may be
higher than the destruction. To examine the later scenario, we estimate the lower limit to the
number density of hydrogen atoms, $n_{H}$, for PAH
survival in the UV photon fields observed:
\begin{equation}
n_{H} > F_{FUV, 5.0 - 13.6 eV}/(H \times c)
\end{equation}
where $c$ is the speed of light, $F_{FUV}$ is the photon flux rate
[$\rm photon~cm^{-2}/sec$] for photons with energy between
5--13.6~eV, and the dimensionless $H$ is defined as the
Habing photodissociation parameter\footnote{``$H$''
represents the ratio of local $F_{FUV}$ photo density and the
density of atomic hydrogen.}, following $\S$12.7 of
\citet{2003adu..book.....D}. \citet{2003adu..book.....D} suggest $H
\sim 0.05$, an upper limit from line-of-sight average from
observation \citep{1996A&A...305..602A}, for PAH molecule
destruction. The Starburst99 code (with Salpeter IMF) gives a ratio
of the far UV ionizing flux, $N_{FUV}$, to $N_{lyc}$ of
$N_{FUV}/N_{lyc} \sim 3$ for O-type dominated stellar population,
thus we find
\begin{equation}\label{eqation:Ffuv}
F_{FUV}[cm^{-2}~sec^{-1}] \sim 1.3 \times
10^{12}~\frac{I_{7mm}}{mJy/bm}.
\end{equation}
So the lower limit of hydrogen atom number density, $n_{H} = 900 \times
(\frac{I_{7mm}}{mJy/bm})~cm^{-3}$ for PAH survival in the localized region 
where $I_{7mm}$ is observed. In M82, the peak 7~mm
VLA flux density in a $\sim 2\arcsec$ beam $\sim 5 - 20~mJy$, requiring $n_{H} >
5 \times 10^3~cm^{-3}$. If $H \sim 0.005$ which ensures the PAH
survival in the PDRs but not inside the ionized media
\citep{2005ApJ...619..755D}, $n_{H} > 5 \times 10^4~cm^{-3}$ is
required to supply the observed PAH, close to the critical density for
HCN excitation. This agrees with the good correlation between intensities of PAH
with HCN emission in M82 found in $2.4\arcsec$ beam (not shown in this paper).

\section{Star Formation History and Efficiency}

\subsection{Star Formation and its Recent History in M82}

\cite{2003ApJ...599..193F} proposed the star formation history of selected
starburst regions in M82 based on ISO mid-IR imaging and spectroscopy
for $\lambda = 1$--$45~\micron$. For the central $16\arcsec \times
10\arcsec$ region, they found star formation had peaked near 5 and 9 Myr ago with
amplitudes of 6.3 and 18.5 $M_{\sun}~yr^{-1}$, respectively, and had
dipped to less than 0.1 $M_{\sun}~yr^{-1}$ at 6 Myr or so. Previously
\citet{2001ApJ...552..544F} had commented on the lack of the
Wolf-Rayet (W-R) spectral signatures in M82, which may be due to high
extinction toward the ``young'' star forming regions (3--6 Myr) where
the W-R stars live, but could also reflect the star formation rate (SFR) dip near 3+ Myr
ago. Our observations suggest that high extinction is the reason 
for the absence of WR features, since all of the young radio free-free 
sources are found in heavily visually obscured regions. 

The integrated flux density at 7 mm of the full region
\cite{2001ApJ...552..544F,2003ApJ...599..193F} mapped is about 340 mJy
in the low resolution 7~mm map. This
equates to a total mass in young stars of $1.3 \times 10^7~M_{\sun}$
formed within the average lifetime of O stars of 1--3 Myr, for a
Salpeter IMF of 1-100 $M_{\sun}$. The contribution of the previous two
starburst events at 5 Myr and 9 Myr to 7~mm flux density would be negligible. The average SFR over this region in about the past
3 Myr (before O stars enter the W-R phase), based on inferred total
mass of young star population from 7~mm flux density, is thus estimated as 4
$M_{\sun}/yr$ (Salpeter IMF of  1-100 $M_{\sun}$), in agreement with the radio recombination line study by
\cite{2004ApJ...616..783R} and with the rate estimated based on
numbers of radio SNR \citep{1994ApJ...424..114H,1994MNRAS.266..455M,2000ApJ...535..706K,2008MNRAS.391.1384F}.

\subsection{Molecular Gas and Star Formation Efficiency}\label{section:SFE}

Our 7~mm continuum images give a picture of where the young star-forming
regions are in M82, particularly the compact 7~mm sources, which are
likely very young embedded clusters. What is the spatial relation of these
sources to the molecular gas? Is there evidence in the spatial distribution
for how the star formation was triggered?

We can estimate $M_{star}$, the mass of new stars formed, from $L_{IR}$, $M_{gas}$, the total mass 
of gas, from $L_{CO}$, and the mass of dense ($M_{dense}$; $n_{H_{2}} >10^{4}~cm^{-3}$) gas from $L_{HCN}$. 
The dense gas fraction is therefore $M_{dense~gas}/M_{gas}$ and the star formation efficiency, the fraction of total gas converted to newly formed stars, is 
$M_{star}/(M_{star}+M_{gas})$.  This is the star formation efficiency for the current generation of stars (i.e., on time scales of a few Myr).
We convert our observed $S_{7mm}$ in lower resolution map to $L_{OB}$ in the 2.7\arcsec\ beam, take that $L_{OB}$ to be  $L_{IR}$, and  find $M_{star}$  from $M_{star}=1.5 \times 10^{-3}~ L_{IR} ~ M_{\sun} ~ L_{\sun}^{-1}$, which assumes 
an O star dominated population with Salpeter IMF between 1-100 $M_{\sun}$. 
Since we are using the total 7~mm flux density for this computation rather than the thermal 7~mm flux density, the $L_{OB}$ and $L_{IR}$ values could be overestimated by 30\% due to the nonthermal contribution.
The full translation is uncertain by a factor of 3, mainly corresponding to the difference in lower mass cutoff in IMF. 
We find the dense gas mass fraction from $I_{HCN}/I_{CO}$. For the molecular gas mass, we adopt the conversion of $M_{dense~gas}(H_{2}) = 10~L_{HCN}~M_{\sun}~(K~km~s^{-1}~pc^{2})^{-1}$, and $M_{gas}(H_{2}) = 4.8~L_{CO}~M_{\sun}~(K~km~s^{-1}~pc^{2})^{-1}$ from \citet{2004ApJS..152...63G}. These conversions for molecular gas are
uncertain by a factor of 2--3, discussed in
\citet{2004ApJS..152...63G}. 

In Figure \ref{figure:Gao_diagram} we plot the observables $L_{IR}/L_{CO}$ vs. $L_{HCN}/L_{CO}$. The corresponding SFE against the 
dense gas fraction (DGF) are shown on the top and right. The data points of M82 represent the results on a 2.7\arcsec\ scale (or $\sim$ 46 pc). We show local normal spirals, 
luminous infrared galaxies (LIRGs), ultra-luminous infrared galaxies (ULIRGs), and 
high-z starburst galaxies from \citet{2007ApJ...660L..93G} spirals, luminous infrared galaxies (LIRGs), ultra-luminous infrared galaxies (ULIRGs), and high-z starburst galaxies from \citet{2007ApJ...660L..93G} on the same plot.  The SFE and DGF track in M82 roughly as they do in all sources, although the size scales of the observations can be very different. All M82 data area lie under the SFE:DGF=1:1 line, which suggests that less than 50\% of the molecular gas has been turned into stars. Most of the data points lie above the SFE:DGF~=~1:10 line, implying a limit of $\sim$ 10\% SFE for dense molecular gas.

\section{Remarks on Compact Non-Thermal Sources}\label{section:nonthermal_compact_rc}

We have detected 5 non-thermal sources with flux densities higher than
5~$\sigma$. Their flux densities at wavelengths between 7~mm and 18~cm in
$0.2\arcsec \times 0.2\arcsec$ beams are listed in Table
\ref{table:flux_compact_200mas} and plotted in Figure
\ref{figure:SED_SNR}. We overlay on the measured SED the best fit to
a model of a single power law with free-free absorption.

The model results are shown in Table \ref{table:compact_em}. About one third of the compact 7~mm sources we detected (Table 1) are
non-thermal. These sources have been known for some time  and are
believed to be SNR \citep{1985Sci...227...28K}. However, they do not
decline with time as expected, so it has been suggested that they
may be wind-driven bubbles instead \citep{2007ApJ...659..347S}.

\subsection{43.992+59.70: Potential AGN in M82?}\label{section:AGN}

The brightest 7~mm continuum compact source is 43.992+59.70, which
has been suggested to be the nucleus of M82 \citep{1975ApJ...200..430K}. 
On the basis of European VLBI Network and VLA images,
\citet{1999MNRAS.305..680W} suggest this source could be a candidate
for the AGN in M82 based on the jet-like feature, although they detected
only the northern part. The study of two radio OH satellite lines also suggests the OH features of this source could be associated with a circumnuclear disk or torus \citep{1997ApJ...487L.131S}. 
43.992+59.70 is also within 3$\arcsec$ of the 2~$\micron$ peak of M82
\citep{1990ApJ...352..544L} that is presumably in the center of the
mass distribution. In addition, [Ne II] line
\citep{1995ApJ...439..163A}, and OH maser kinematics
\citep{1997ApJ...487L.131S}, molecular gas investigation
\citep{1997ApJ...487L.131S}, and radio recombination line results
\citep{2004ApJ...616..783R} also suggest that this source is close
to the dynamical center of M82, which supports the AGN scenario.

The suggestion that 43.992+59.70 is an AGN can be also tested by looking
for variability. Long-term radio flux monitoring by
\citet{2000ApJ...535..706K} suggests that the source has a
stable flux density, with variability $\leq$ 5\%, which is the uncertainty  
in the absolute flux scale calibration. Our
7~mm flux density is about 63\% of the archival centimeter-wave SED
model's prediction. It is possible that 43.992+59.70 is variable at the
current epoch, or that the spectrum is not power law to 7~mm.  

A discrepancy from the AGN picture is that Chandra high resolution
X-ray images do not show 43.992+59.70 as a strong source. Also, another argument against the AGN nature of 43.992+59.70 can be based on
the recent high resolution VLBI and MERLIN images of this source \citep{2001MNRAS.322..100M,2008MNRAS.391.1384F}, where it has a resolved round
structure, consistent with an SNR as suggested by its spectral index ($\alpha \sim -0.5$) 
\citep{1994ApJ...424..114H}. However, the measured 7~mm flux density appears to miss a significant amount of flux density from the SED model of SNR, and it is hard to match the 7~mm flux density into the SNR picture.

This bright non-thermal source also coincides with a compact
$Pa\,{\alpha}$ and 1.644 \micron~[Fe\,{\sc ii}] emitter.
\cite{2007ApJ...659..347S} modeled the radio emission of non-thermal
sources in M~82, including this one, as wind-driven bubbles. This
scenario hypothesizes that non-thermal synchrotron emitting
particles are produced at the shock between the wind of young star
clusters and hot bubble gas. 
However, \citet{2008MNRAS.391.1384F} measure the expension velocity of this source to be at 2700 km/s, which is about 50 times higher than the expect wind-driven shell \citep{2007ApJ...659..347S}, and it is more closer to the free-expending SNR.

In summary, these three models involving AGN, SNR, and non-thermal bubble of a young cluster
could all explain the strong 7~mm emission of 43.992+59.70. 
Despite some suggestive dynamical evidences, the morphology and expansion revealed by the VLBI and MERLIN work strongly indicate that the source is an SNR. However, the SNR scenario could not explain the decline of flux density at 7~mm from the expectation value of SED modeling. The nature of this source might be still open to debate.

\subsection{Intermediate Black Hole Candidate M82 X-1 (CXO M82
J095550.2+694047)}

\cite{2001ApJ...547L..25M} reported 9 X-ray sources in M82. The
source CXO M82 J095550.2+694047, or M82 X-1
\citep[$09^{h}55^{m}50.2^{s}$,
$+69\degr40\arcmin47\arcsec$;][]{2001ApJ...547L..25M}, has been
categorized as an ultraluminous X-ray source (ULX) and a 54 mHz
quasi-periodic oscillation has been found in X-ray by
\citet{2003ApJ...586L..61S} and a possible 62 day X-ray luminosity
variation by \citet{2006ApJ...646..174K}. It is suggested to host an
intermediate mass black hole \citep{2001ApJ...562L..19E,
2004Natur.428..724P}. Many attempts have been made to find a radio
counterpart of M82 X-1 \citep{2001ApJ...547L..25M,
2005A&A...436..427K, 2006ApJ...646..174K}. The source 41.30+59.6 or
41.5+59.7 has been suggested to be a possible counterpart. We did
not detect a significant radio counterpart of M82 X-1 at 7~mm to a
limit of 0.7 mJy/bm. The SEDs of the nearby 7~mm sources
40.933+58.98 and 41.628+58.01 are clearly thermal and are also not
candidates for the radio counterpart of M82 X-1.

\subsection{Variable Compact
Sources}\label{section:individual_nonthermal}

The flux density of nonthermal source 41.942+57.62 (41.9+58 in
\citealt{1985ApJ...291..693K} and \citealt{2000ApJ...535..706K}) at
7~mm is significantly below the power law model (Fig.
\ref{figure:SED_SNR}). This source is known to vary
\citep{1985Sci...227...28K}; 
its flux density declines at $\sim$ 8.4 -- 8.8\% per year at $\sim$ 5~GHz,
as has been shown and discussed by several authors \citep{1992xrea.conf..247K,1999PhDT........10A,2000ApJ...535..706K,2005MmSAI..76..586M,2008MNRAS.391.1384F}.
The cm data listed in Table \ref{table:flux_compact_200mas} and plotted in
Figure \ref{figure:SED_SNR} were obtained before 1996 \citep{1999PhDT........10A}. 
Based on the epoch-corrected 1993 values of \citet{1998ApJ...502..218A} for the flux density of 41.942+57.62, and for a spectral index of -0.72 (from best-fit result without 7~mm data point), we estimate that the flux density of 41.942+57.62 would have been 8 mJy at 7~mm in 1993, three times higher than the observed value. This is consistent with a decline of 8.4\%/year found at 5 GHz.

\citet{2004IAUC.8297....2S} first reported the supernova SN2004am
(9:55:46.61; +69:40:48.1), which was classified spectroscopically as
Type II \citep{2004IAUC.8299....2M}. \citet{2004IAUC.8332....2B}
report 3-$\sigma$ upper limits for this source of 0.34 mJy/bm (8.4
GHz) and 1 mJy/bm (15 GHz) on 2003 November 14. The null detection
in their MERLIN observations at 5 GHz on March 9 2004 gave an upper
limit of 0.18 mJy/bm (3 $\sigma$), and suggested that the SN2004am
should reach the radio emission peak with a significant delay (up to
about 1 year). We can place a 3~$\sigma$ upper limit of
$S_{7mm}=0.6$ mJy/bm to the 7~mm flux density of SN2004am in our observation
of 2005 April 03.

There are at least 7 known sources for which the 7~mm flux density predicted
from their cm-wave flux densities should be larger than 1~mJy ($5 - 7
\sigma$). They are 40.68+551, 43.82+628, 45.41+637, 45.48+647,
46.18+677, 46.52+639, and 47.37+680.  The last four of these are at
$>20\arcsec$ from the phase center where the noise level is a factor
of 2 higher than elsewhere and these may be down in the noise.
Sources 40.68+551 and 43.82+628 are, however, within the central
primary beam in a low-noise region. Both 40.68+551 and 43.82+628 are
SNR. Source 40.68+551 has minor variability in the long term radio
continuum flux experiment of \citet{2000ApJ...535..706K}, but no variability of 43.82+628 has been reported, and this may be
the first indication.

For three of the 7~mm sources (39.659+55.73, 43.992+59.70, and
45.153+61.34) the 7~mm flux densities are stronger at the epoch of our
observation (2005) with $\lesssim 0.2\arcsec$ beam than at year
2000-2001 in the observation of \citet{2004ApJ...616..783R} with a
$0.6\arcsec$ beam. The flux density differences are not conclusive ($\sim 2~\sigma$ 
of the C-array), but suggest that these sources may be
varying. The latter two sources are believed to be SNR, based on
their radio SEDs. Neither of them vary strongly at 6~cm in the
12-year monitoring program of \citet{2000ApJ...535..706K}  between
1981--1992. Source 39.659+55.73 is known to vary at 6~cm,
fluctuating 20\% continuously during this time period. But we have
characterized this source as an \HII\ region based on its rising
spectrum, and it therefore should not vary. This source might be
associated with a wind-driven bubble, as suggested for the
non-thermal sources by \citet{2007ApJ...659..347S}.

\section{Conclusions}

We present 7~mm VLA continuum maps of the starburst in M82 at two resolutions, 0.2\arcsec\ (B-configuration) and 2\arcsec\ (D-configuration). We recover essentially all of the 7~mm flux density of M82 in the lower resolution D-array map. An estimated 2/3 of this flux density is thermal free-free emission from the starburst. We detect less than 10\% of the total 7~mm flux density in the high resolution B-array map, which is sensitive to only bright, compact emission sources. 

We identify 14 compact sources greater than 0.5~mJy/bm (5$\sigma$) in the high resolution image, and have constructed SEDs using archival VLA data. Nine of the compact sources are compact \HII\ regions, with emission measures $EM>10^8~\rm cm^{-6}\,pc,$ sizes of $\lesssim 3$~ pc, and rms densities $<n_e^2>^{1/2}\gtrsim 3000$. These compact sources are located in regions of heavy visual obscuration, with no associated near-IR continuum counterparts. This could explain the lack of W-R features in M82.
Our detection limit is $N_{lyc} \sim 0.6 \times 10^{51}~\rm s^{-1}$, so the smallest of these clusters has $\sim 100$ O stars, and the largest, $\sim$ 1200 O stars ($N_{lyc} \sim 10^{53}~sec^{-1}$). The detected compact thermal 
sources account for $\sim$ 6\%\ of the total 7~mm free-free flux density. 
Since the maps detect only the most luminous compact \HII\ regions, this ratio should be higher if fainter ones are taken into account. We therefore infer that the compact \HII\ region phase in M82 must last at least 6\%\ of the \HII\ region lifetime, $\sim$1~Myr or longer. 

In addition to the compact \HII\ regions, the high resolution map reveals extended (2\arcsec, or 30 pc) regions of free-free emission with $EM>10^7~\rm cm^{-6}\,pc$. 
The comparison of intensity (EM) and flux ($N_{Lyc}$) suggests that these regions are 
likely to be non-spherical in geometry. We suggest two possibilities, either these regions indicate regions of distributed star formation, with many small \HII\ regions possibly reflecting enhanced star formation at the ends of the bar in M82, or these regions are sheet-like structures, possibly  caused by winds or supernovae.

We have compared lower resolution 7~mm VLA maps to maps of molecular gas tracers and dust. The good correlation between the 7~mm emission and \Spitzer\ IRAC Channel 4 map (dominated by $8~\micron$ PAH features) suggests a relation $N_{Lyc} \sim 2.2 \times 10^{51}~s^{-1} (\frac{D}{Mpc})^{2} (\frac{S_{8\mu m}}{Jy})$, for intensities $I_{8\micron} > 0.06~mJy/pc^{2}$. This indicates that the PAH feature is a good tracer of star formation in M82, although we obtain a different relation between $N_{Lyc}$ and the $8~\micron$ emission than previous work. 

The correlation of 7~mm emission and molecular gas tracers (CO and HCN) at 2\arcsec\ scales is excellent, indicating that the relation of star formation and HCN emission found on global scales by \citep{2004ApJ...606..271G} holds down to GMC scales. We estimate that the SFE in M82 on GMC scales is $\sim$ 1--10\%, similar to Galactic values, and that the regions of highest SFE also have the highest fraction of dense ($n_{H_{2}}\gtrsim 10^{4}~cm^{-3}$) gas.

\acknowledgments

The authors would like to thank Almudena Alonso-Herrero, Sarah Lipscy, 
and Peter Schilke for their generosity in providing their data.
C.W.T. appreciates helpful discussions with Yu Gao. Some of the data
used in this paper were obtained from the Multimission Archive at
the Space Telescope Science Institute (MAST). STScI is operated by
the Association of Universities for Research in Astronomy, Inc.,
under NASA contract NAS5-26555. This research has also made use of
the NASA/IPAC Extragalactic Database (NED) and the NASA/ IPAC
Infrared Science Archive (IRSA) which are operated by the Jet
Propulsion Laboratory, California Institute of Technology, under
contract with the National Aeronautics and Space Administration.

\clearpage

%%%%%%%%%%%%%%%%%FIGURES%%%%%%%%%%%%%%%%%%%%

%%%%%%%%%%%%%%%%%%%TABLES%%%%%%%%%%%%%%%%%%%%%%
\clearpage
\begin{deluxetable}{llccccl} 
\tabletypesize{\scriptsize}
\tablewidth{0in}
\tablecaption{Compact 7~mm Continuum Sources: Positions and Fluxes\label{table:flux_compact}}
\tablehead{
\multicolumn{1}{c}{Source\tablenotemark{a}} &
\multicolumn{1}{l}{Other Name\tablenotemark{b}} &
\multicolumn{1}{c}{$\alpha$(J2000)\tablenotemark{c}} &
\multicolumn{1}{c}{$\delta$(J2000)\tablenotemark{c}} &
\multicolumn{1}{c}{$S_{7 mm}^{Peak}$\tablenotemark{;d}} &
\multicolumn{1}{c}{$S_{7 mm}$\tablenotemark{d}} &
\multicolumn{1}{c}{Size\tablenotemark{e}}
\\
\colhead{(B1950)} &
\colhead{(A99/M02)} &
\colhead{09$^{h}$55$^{m}$+ ($^{s}$)} &
\colhead{69$\arcdeg$40$\arcmin$+ ($\arcsec$)} &
\colhead{(mJy beam$^{-1})$} &
\colhead{(mJy)} &
\colhead{$\arcsec$ $\times$ $\arcsec$, $\degr$}}
\startdata
 39.277+54.32 & 39.29+54.2 &   48.051  &  40.64  & 0.9 & 1.4$\pm$0.4 & 0.31, 0.25, 60 \\ 
 39.659+55.73 & 39.68+55.6 &   48.427  &  41.97  & 3.6 & 4.6$\pm$0.3 & 0.21, 0.18, 170 \\ 
 40.933+58.98 & 40.94+58.8 &   49.694  &  45.20  & 0.6 & 2.3$\pm$0.6 & 0.47, 0.35, 70  \\ 
 41.145+56.29 & 41.17+56.2 &   49.904  &  42.51  & 0.8 & 3.1$\pm$0.6 & 0.38, 0.30, 170 \\ 
 41.628+58.01 & 41.64+58.4 &   50.375  &  44.22  & 1.2 & 3.5$\pm$0.5 & 0.33, 0.26, 0   \\ 
 41.942+57.62 & 41.95+57.5 &   50.686  &  43.77  & 2.6 & 2.1$\pm$0.2 & 0.18, 0.13, 180 \\ 
 42.072+58.52 & 42.58+58.4 &   50.817  &  44.69  & 1.2 & 2.2$\pm$0.3 & 0.28, 0.19, 10  \\ 
 42.193+59.13 & 42.21+59.2 &   50.945  &  45.32  & 1.4 & 2.6$\pm$0.3 & 0.26, 0.21, 130 \\ 
 42.549+58.13 & 42.56+58.0 &   51.291  &  44.29  & 0.9 & 9.2$\pm$1.3 & 0.68, 0.44, 140\tablenotemark{f} \\ 
 42.673+58.32 & 42.69+58.2 &   51.414  &  44.47  & 0.6 & 9.3$\pm$1.5 & 0.82, 0.52, 140\tablenotemark{f} \\ 
 43.173+58.43 & 43.18+58.3 &   51.908  &  44.56  & 0.7 & 1.2$\pm$0.3 & 0.25, 0.19, 90  \\ 
 43.295+59.27 & 43.31+59.2 &   52.028  &  45.42  & 1.6 & 2.0$\pm$0.3 & 0.22, 0.17, 170 \\ 
 43.992+59.70 & 44.01+59.6 &   52.723  &  45.78  & 5.2 & 5.4$\pm$0.2 & 0.20, 0.16, 180 \\ 
 45.153+61.34 & 45.17+61.2 &   53.875  &  47.41  & 1.1 & 1.9$\pm$0.3 & 0.23, 0.21, 170  
\enddata
\tablenotetext{a}{Convention of Kronberg et al. (1985).}
\tablenotetext{b}{A99: \citet{1999PhDT........10A}; M02: \citet{2002MNRAS.334..912M}.}
\tablenotetext{c}{The typical uncertainty of position 
within the 7~mm map is $\sim$ 10 -- 20 mas for strong sources, and 30 -- 40 mas for weaker
sources (flux density $\lesssim$ 1 mJy/bm).
The absolute astrometry accuracy is estimated to be $\sim 0.1\arcsec$, limited by the atmospheric phase stability during the observation.
}
\tablenotetext{d}{$S_{7~mm}^{Peak}$ and $S_{7~mm}$: peak flux density and total integrated flux density of 7~mm continuum,
respectively, with beam as $0.19\arcsec \times 0.15\arcsec$, PA= -10$\degr$. The rms uncertainty of each measurement is 0.13~mJy/bm. The systematic error of flux density is expected to be $\lesssim$ 5\%.}.
\tablenotetext{e}{Convolved size. Single Gaussian is assumed for each source; using {\sc aips} task {\sc imfit}.
The typical uncertainty of source size (convolved) is $\lesssim$ 10\% except source 42.69+58.2 and 42.56+58.0 in which the uncertainty is $\sim$ 20\%.}
\tablenotetext{f}{Peaks in emission complex.}
\end{deluxetable}

\begin{deluxetable}{crrrrrrl} 
\tabletypesize{\scriptsize}
\tablewidth{0in}
\tablecaption{Radio SEDs of Compact Sources\label{table:flux_compact_200mas}}
\tablehead{
\multicolumn{1}{c}{Source\tablenotemark{a}} &
\multicolumn{1}{r}{$S_{18 cm}$\tablenotemark{b}} &
\multicolumn{1}{r}{$S_{6 cm}$\tablenotemark{b}} &
\multicolumn{1}{r}{$S_{3.6 cm}$\tablenotemark{b}} &
\multicolumn{1}{r}{$S_{2 cm}$\tablenotemark{b}} &
\multicolumn{1}{r}{$S_{1.3 cm}$\tablenotemark{b}} &
\multicolumn{1}{r}{$S_{7 mm}$\tablenotemark{b}} &
\multicolumn{1}{c}{Reference\tablenotemark{c}} \\
\colhead{(B1950)} &
\colhead{(mJy)} &
\colhead{(mJy)} &
\colhead{(mJy)} &
\colhead{(mJy)} &
\colhead{(mJy)} &
\colhead{(mJy)}}
\startdata
 39.277+54.32 &  \dots          & 1.7  $\pm$ 0.2 & \dots          & 1.5  $\pm$ 0.2 & \dots           & 1.5 $\pm$ 0.5 & M02\\ 
 39.659+55.73 &  \dots          & 1.5  $\pm$ 0.2 & \dots          & 3.2  $\pm$ 0.2 & \dots           & 8.0 $\pm$ 0.6 & M02 \\ 
 40.933+58.98 &  \dots          & 1.9  $\pm$ 0.2 & 2.9 $\pm$ 0.1 & 2.8 $\pm$ 0.3 & 2.2 $\pm$ 0.3 & 2.5 $\pm$ 0.4  & A99\\ 
 41.145+56.29 &  \dots          & 2.0  $\pm$ 0.3 & \dots          & 2.8  $\pm$ 0.2 & \dots           & 5.1 $\pm$ 0.6  & M02\\ 
 41.628+58.01 &  \dots          & 1.0  $\pm$ 0.2 & \dots          & 1.8  $\pm$ 0.2 & \dots           & 6.1 $\pm$ 0.5  & M02\\ 
 41.942+57.62\tablenotemark{d} &  79.0 $\pm$ 4.0 & 40.0 $\pm$ 0.9 & 26.8 $\pm$ 0.5 & 17.9 $\pm$ 1.7 & 12.0 $\pm$ 0.6  & 2.4 $\pm$ 0.3  & AK98\\ 
 42.072+58.52 &  \dots          & 1.2  $\pm$ 0.4 & \dots          & 1.7  $\pm$ 0.2 & \dots           & 3.6 $\pm$ 0.6  & M02\\ 
 42.193+59.13 &  2.5  $\pm$ 1.0 & 4.1  $\pm$ 0.5 & 4.3  $\pm$ 0.3 & 4.5  $\pm$ 0.3 & 4.1  $\pm$ 0.3  & 4.0 $\pm$ 0.5  & AK98\\ 
 42.549+58.13 &  \dots          & 2.7  $\pm$ 0.4 & \dots         & 1.8 $\pm$ 0.2 & \dots          & 1.7 $\pm$ 0.2\tablenotemark{e}  & M02\\ 
 42.673+58.32 &  \dots          & 1.0  $\pm$ 0.4 & \dots         & 1.8 $\pm$ 0.2 & \dots          & 1.1 $\pm$ 0.2\tablenotemark{e}  & M02\\ 
 43.173+58.43 &   8.5 $\pm$ 1.0 & 5.2  $\pm$ 0.2 & 3.8 $\pm$ 0.2 & 2.4 $\pm$ 0.3 & 1.8 $\pm$ 0.3  & 1.6 $\pm$ 0.5  & AK98\\ 
 43.295+59.27 &  19.5 $\pm$ 1.0 & 10.3 $\pm$ 0.2 & 8.0  $\pm$ 0.1 & 5.5  $\pm$ 0.3 & 4.1  $\pm$ 0.4  & 2.5 $\pm$ 0.5  & AK98\\ 
 43.992+59.70 &  9.5  $\pm$ 1.0 & 21.8 $\pm$ 0.9 & 20.2 $\pm$ 0.1 & 15.8 $\pm$ 0.3 & 12.5 $\pm$ 0.3  & 5.4 $\pm$ 0.4  & AK98\\ 
 45.153+61.34 &  12.5 $\pm$ 1.0 & 7.9  $\pm$ 0.3 & 5.7  $\pm$ 0.1 & 3.9  $\pm$ 0.3 & 2.5  $\pm$ 0.3  & 2.2 $\pm$ 0.5 & AK98
 
\enddata
\tablenotetext{a}{Convention of Kronberg et al. (1985)}
\tablenotetext{b}{$S_{\lambda}$: Total integrated flux density of radio continuum at wavelength $\lambda$ with 0.2$\arcsec$ to 0.3$\arcsec$ beam. See \S\ref{section:overview} for details.}
\tablenotetext{c}{The flux densities at 18 -- 1.3~cm are adopted from AK98: Allen \& Kronberg (1998; and references therein); A99: Allen et al (1999); M02: McDonald et al. (2002).}
\tablenotetext{d}{The source is a known strong variable. The flux density densities measured at different epochs have been corrected to the epoch of 1993.4 using a decline rate of 8.8\% \citep[see][for details]{1998ApJ...502..218A}, except 7~mm flux density, which is measured in 2005 in this work.}
\tablenotetext{e}{Source is in emission complex. The peak flux density at
7~mm obtained with {\sc imfit}
is used. This could be an underestimate if the source is resolved.}
\end{deluxetable}

\begin{deluxetable}{cclcrrrrc} 
\tabletypesize{\scriptsize}
\tablewidth{0in} \tablecaption{SED Models of Compact
Sources\label{table:compact_em}} \tablehead{
\multicolumn{1}{c}{Source} &
\multicolumn{1}{c}{Type\tablenotemark{a}} &
\multicolumn{1}{c}{Model\tablenotemark{b}} &
\multicolumn{1}{c}{$\alpha$\tablenotemark{c}} &
\multicolumn{1}{c}{$\nu_{\tau = 1}$\tablenotemark{c}} &
\multicolumn{1}{c}{EM\tablenotemark{d}} &
\multicolumn{1}{c}{$N_{Lyc}\tablenotemark{e}$} &
\multicolumn{1}{c}{$N_{O}$} &
\multicolumn{1}{c}{$L_{OB}$}
\\
\colhead{(B1950)} &
\colhead{} &
\colhead{} &
\colhead{} &
\colhead{(GHz)} &
\colhead{($10^6~pc~cm^{-6}$)} &
\colhead{$10^{50}~sec^{-1}$} &
\colhead{} &
\colhead{$10^{8}~L_{\sun}$} }
\startdata
 39.277+54.32 & \HII\ & PL     & -0.1  & $<5$  & $<90$   & 22  & 220  & 0.4 \\
 39.659+55.73 & \HII\ & PL     &  0.8  & $>43$ & $>8000$ & 116 & 1160 & 2.3 \\
 40.933+58.98 & \HII\ & FFSA   & \dots & 4.9   & $85$    & 36  & 360  & 0.7 \\
 41.145+56.29 & \HII\ & PL     &  0.5  & $>43$ & $>8000$ & 74  & 740  & 1.5 \\
 41.628+58.01 & \HII\ & PL     &  1.0  & $>43$ & $>8000$ & 88  & 880  & 1.8 \\
 42.072+58.52 & \HII\ & PL     &  0.6  & $>43$ & $>8000$ & 52  & 520  & 1.0 \\
 42.193+59.13 & \HII\ & FFSA   & \dots & 2.5   & $20$    & 58  & 580  & 1.2 \\
 42.549+58.13 & \HII\ & PL     & -0.2  & $<5$  & $<90$   & 25  & 250  & 0.5 \\
 42.673+58.32 & \HII\ & PL     & -0.1  & $<5$  & $<90$   & 16  & 160  & 0.3 \\
 \tableline
 41.942+57.62\tablenotemark{f} &  SNR & PL+FFA & -0.7/-0.9\tablenotemark{f}  & \dots & \dots  \\
 43.173+58.43 &  SNR & PL+FFA  & -0.7  & 0.9   & $2.7$  \\
 43.295+59.27 &  SNR & PL+FFA  & -0.6  & \dots & \dots  \\
 43.992+59.70 &  SNR & PL+FFA  & -0.6  & 2.2   & 6.6    \\
 45.153+61.34 &  SNR & PL+FFA  & -0.7  & 1.0   & 3.0
\enddata
\tablenotetext{a}{Based on the previous cm results from \cite{1998ApJ...502..218A,1999PhDT........10A,2002MNRAS.334..912M}.}
\tablenotetext{b}{PL: single power law; FFSA: free-free emission with self-absorption; FFA: foreground free-free absorption.}
\tablenotetext{c}{$\alpha$: spectral index, $S_{\nu} \propto \nu^{\alpha}$. $\nu_{\tau}$: turnover frequency.}
\tablenotetext{d}{$EM \propto (\nu_{\tau = 1})^{2.1}$. $T_{e} = 10,000$~K is assumed.}
\tablenotetext{e}{Assuming optically thin free-free emission. For optically thick sources, the numbers would be the lower limits.}
\tablenotetext{f}{Variable source. $\alpha = -0.7$: with epoch corrected (1993.4) data at wavelengths at 18 -- 1.3~cm. $\alpha = -0.9$: with all radio continuum data, including flux density at 7~mm of epoch 2005. See \S \ref{section:individual_nonthermal}.}
\end{deluxetable}

        \clearpage
%        \LongTables % optionally
        \begin{landscape}
\begin{deluxetable*}{lcccccccccc} 
%\begin{deluxetable}{lcccccccccc} 
%\rotate 
\tabletypesize{\scriptsize} \tablewidth{0pt}
\tablecaption{Properties of Peaks in Low Resolution
Map\label{table:peaks_Darray}} \tablehead{
\multicolumn{1}{l}{Region\tablenotemark{a}} &
\multicolumn{1}{c}{Peak Position\tablenotemark{b}} &
\multicolumn{2}{c}{\hrulefill~~$S_{7 cm}$\tablenotemark{c}~\hrulefill} &
\multicolumn{1}{c}{EM\tablenotemark{d}} & 
\multicolumn{3}{c}{\hrulefill~~$N_{Lyc}~~$\hrulefill} &
\multicolumn{1}{c}{$N_{O7}$\tablenotemark{d}} &
\multicolumn{1}{c}{$L_{OB}$\tablenotemark{d}} &
\multicolumn{1}{c}{$L_{MIR}$\tablenotemark{f}}
\\
\colhead{}
&\colhead{(J2000)}
&\colhead{Peak}
&\colhead{Total}
&\colhead{}
&\colhead{(7~mm\tablenotemark{d})}
&\colhead{(100~GHz\tablenotemark{e})}
&\colhead{(MIR\tablenotemark{f})}
\\
\colhead{}
& \colhead{(09$^{h}$55$^{m}$+; 69$\arcdeg$40$\arcmin$+)}
& \colhead{(mJy/bm)}
& \colhead{(mJy)}
& \colhead{($10^{6}~pc~cm^{-6}$)}
& \colhead{($10^{52}~sec^{-1}$)}
& \colhead{($10^{52}~sec^{-1}$)}
& \colhead{($10^{52}~sec^{-1}$)}
& \colhead{($\times$100)}
& \colhead{($10^{8}~L_{\sun}$)}
& \colhead{($10^{8}~L_{\sun}$)}
 }
\startdata
W1          & 48.08, 40.8 & 2.4 &  4 & 16.0 &  0.6 & \dots & \dots &   6 &  1.2 & \dots \\
W2          & 48.44, 42.0 & 4.1 &  8 & 10.5 &  1.2 & \dots & \dots &  12 &  2.3 & \dots \\
W3          & 49.77, 44.3 & 8.2 & 39 & 12.5 &  5.7 & \dots & 0.08  &  57 & 11 & 0.2   \\
W4          & 49.96, 42.9 & 8.1 & 38 & 12.9 &  5.5 & 2.8   & \dots &  55 & 11 & \dots \\
W4a         & 49.50, 42.3 & 4.6 & 19 &  7.4 &  2.8 & \dots & \dots &  28 &  5.5 & \dots \\
W5          & 50.29, 44.1 &  11 & 65 & 15.5 &  9.4 & 3.6   & 2.4   &  94 & 19 & 5.9   \\
R1          & 50.98, 45.2 &  13 & 81 & 18.5 & 11.8 & 3.8   & 2.2   & 12 & 24 & 5.5   \\
R2          & 51.33, 44.4 &  15 & 82 & 22.3 & 11.9 & 4.2   & 1.3   & 118 & 24 & 3.3   \\
R3          & 51.30, 46.2 & 9.5 & 65 & 13.0 &  9.4 & \dots & \dots &  94 & 19 & \dots \\
R4          & 52.06, 48.1 & 6.9 & 41 & 10.4 &  6.0 & 2.3   & 0.7   &  60 & 12 & 1.7   \\
R5          & 52.55, 48.8 & 7.3 & 40 & 10.6 &  5.8 & \dots & 0.9   &  58 & 12 & 2.3   \\
R6          & 52.77, 46.2 & 6.0 & 23 &  9.5 &  3.3 & \dots & \dots &  33 &  6.7 & \dots \\
R7          & 53.06, 48.1 & 5.9 & 39 &  8.4 &  5.7 & 2.2   & 2.0   &  57 & 11 & 4.9   \\
E1          & 53.66, 49.7 & 4.3 & 29 &  6.1 &  4.2 & \dots & \dots &  42 &  8.4 & \dots \\
E1a         & 53.86, 48.1 & 3.4 & 10 &  6.4 &  1.5 & \dots & \dots &  15 &  2.9 & \dots \\
E2          & 54.00, 51.4 & 8.1 & 54 & 11.5 &  7.8 & 2.4   & \dots &  78 & 16 & \dots \\
E3          & 54.82, 53.4 & 2.7 & 16 &  3.8 &  2.3 & \dots & \dots &  23 &  4.6 & \dots
\enddata
\tablenotetext{a}{Following the name in Allen (1999) except W4a and E1a which are newly identified in this work.}
\tablenotetext{b}{Typical peak position uncertainty is $\sim 0.1\arcsec - 0.2\arcsec$.}
\tablenotetext{c}{In robust weighted, VLA D-array map at 45 GHz with $1.6\arcsec \times 1.2\arcsec$, PA = $20\degr$ beam, rms = 0.3 mJy/bm; using 
{\sc imfit} task in {\sc aips}}.
\tablenotetext{d}{Assume $T_{e} = 10,000~K$, and pure optically thin thermal emission at 7~mm. The possible non-thermal contamination is estimated to be $\lesssim 35\%$.}
\tablenotetext{e}{Adopted from \cite{2005ApJ...618..712M}. Derived from peak flux density at 100~GHz with $2.3\arcsec \times 1.9\arcsec$ beam. Assume $T_{e} = 10,000~K$.}
\tablenotetext{f}{Adopted from \cite{2004ApJ...603...82L}. Their seeing-limited MIR images are in $0.4\arcsec - 1.0\arcsec$ resolution.}
%\end{deluxetable}
\end{deluxetable*}
\clearpage
        \end{landscape}

\end{document}